\documentclass{article}

\usepackage{arxiv}

\usepackage[utf8]{inputenc} 
\usepackage[T1]{fontenc}    
\usepackage{hyperref}       
\usepackage{url}            
\usepackage{booktabs}       
\usepackage{amsfonts}       
\usepackage{nicefrac}       
\usepackage{microtype}      
\usepackage{lipsum}
\usepackage{graphicx}
\graphicspath{ {./images/} }

\usepackage{graphicx}
\usepackage{newtxtext}
\usepackage{newtxmath}
\usepackage{natbib}
\usepackage{hyperref}
\usepackage{tikz}
\usepackage{overpic}
\usepackage{transparent}
\usepackage{adjustbox} 
\usepackage{gensymb}
\usepackage{soul}
\newcommand{\mathsfbi}[1]{\mathbf{\mathsf{#1}}}
\newcommand{\norm}[1]{\left\lVert#1\right\rVert}

\usetikzlibrary{arrows, positioning, calc}
\hypersetup{
    colorlinks = true,
    urlcolor   = blue,
    citecolor  = black,
}

\newcommand{\RomanNumeralCaps}[1]
\linenumbers

\usepackage{bm} 

\tikzset{
    block/.style={draw, rectangle, minimum height=2em, minimum width=4em},
    sum/.style={draw, circle, inner sep=0pt, minimum size=4mm},
    delay/.style={draw, rectangle, minimum height=2em, minimum width=4em},
    input/.style={coordinate},
    output/.style={coordinate}
}

\title{Observer-based neural networks for flow estimation and control}

\author{
 Tarcísio C. Déda \\
  School of Mechanical Engineering\\
  University of Campinas\\
  R. Mendeleyev 200, Campinas, São Paulo, Brazil \\
  \texttt{tdeda@unicamp.br} \\
   \And
 William R. Wolf \\
  School of Mechanical Engineering\\
  University of Campinas\\
  R. Mendeleyev 200, Campinas, São Paulo, Brazil \\
  \texttt{wolf@fem.unicamp.br} \\
  \And
 Scott T. M. Dawson \\
  Mechanical, Materials, and Aerospace Engineering\\
  Illinois Institute of Technology\\
  10 W 35th St, Chicago, IL, United States \\
  \texttt{sdawson5@iit.edu} \\
   \And
 Brener L. O. Ramos \\
  School of Mechanical Engineering\\
  University of Campinas\\
  R. Mendeleyev 200, Campinas, São Paulo, Brazil \\
  \texttt{brener.lelis@gmail.com} \\
}

\begin{document}
\maketitle
\begin{abstract}
Neural network observers (NNOs) are proposed for real-time estimation of fluid flows, addressing a key challenge in flow control: obtaining real-time flow states from a limited set of sparse and noisy sensor data.
For this task, we propose a generalization of the classical Luenberger observer. In the present framework, the estimation loop is composed of subsystems modeled as neural networks (NNs). By combining flow information from selected probes and an NN surrogate model (NNSM) of the flow system, we train NNOs capable of fusing information to provide the best estimation of the states, that can in turn be fed back to an NN controller (NNC). The NNO capabilities are demonstrated for three nonlinear dynamical systems. First, a variation of the Kuramoto-Sivashinsky (KS) equation with control inputs is studied, where variables are sparsely probed. We show that the NNO is able to track states even when probes are contaminated with random noise or with sensors at insufficient sample rates to match the control time step. Then, a confined cylinder flow is investigated, where velocity signals along the cylinder wake are estimated by using a small set of wall pressure sensors. In both the KS and cylinder problems, we show that the estimated states can be used to enable closed-loop control, taking advantage of stabilizing NNCs. Finally, we present a legacy dataset of a turbulent boundary layer experiment, where convolutional NNs (CNNs) are employed to implement the models required for the estimation loop. We show that, by combining low-resolution noise-corrupted sensor data with an imperfect NNSM, it is possible to produce more accurate estimates, outperforming both the direct reconstructions via specialized super-resolution NNs and the direct model propagation from initial conditions. 
\end{abstract}


\section{Introduction}

Active control of fluidic systems is a challenging task, partly due to complex nonlinear phenomena and high dimensionality. Closed-loop flow control requires particular attention to challenges related to sensor placement, data acquisition systems, real-time adjustable actuators, and sampling specifications (e.g., sensor bandwidth, noise levels, and sampling frequency), which are often limited by cost and technological limitations. Frameworks commonly studied in flow sciences, such as resolvent analysis, machine learning, and network analysis often serve as effective tools for designing control algorithms \citep{gadelhak2001future,brunton2015closed,brunton2020machine,taira2022network,audiffred2023experimental}.

Resolvent analysis has been combined with experimental setups to model fluidic systems and design optimal controllers.
For instance, \cite{audiffred2023experimental} achieved the attenuation of flow unsteadiness using plasma actuators to cancel incoming Tollmien-Schlichting waves. Although their setup did not involve feedback control, a feedforward resolvent-based methodology was proposed to find an optimal causal control kernel via the Wiener-Hopf formalism. The experimental results showed that the approach is capable of promoting better attenuations compared to direct cancellation through transfer function inversion, which requires truncation to a causal kernel that leads to suboptimal solutions. 

The development of learning techniques is also leveraged in the field of experimental flow control. \cite{fan2020reinforcement} developed a closed-loop control system using reinforcement learning to maximize power gain efficiency in a cylinder flow by evaluating the states that comprise the drag and lift coefficients. Their actuation setup used smaller rotating cylinders in the wake, enabling considerable drag reduction.

In recent studies, researchers explored new control concepts through numerical simulations, which provide controlled environments with access to variables that are often inaccessible in experiments. \cite{rabault2019artificial}, \cite{ren2021applying}, \cite{castellanos2022machine},\cite{varela2022deep} and \cite{deda2024neural} employed machine learning techniques to control the flow past a confined cylinder and successfully achieved performance goals such as stabilization and drag reduction. One of the setups used in these studies---which is also explored in the current work---makes use of minijets actuating in phase opposition, modulated by feedback of flow field velocity measurements. The machine learning algorithms proposed in these studies were shown to be powerful tools for controlling idealised flow systems. Closed-loop control, however, relied on feedback of wake velocity measurements, and the impracticality of such sensing approach motivates research on flow estimation via more realistic real-time sampling of variables.

Control of turbulent flows within numerical environments have also been studied \citep{guastoni2023deep,park2020machine}. \cite{sonoda2023reinforcement} presented a neural network (NN) approach to perform continuous actuation along the walls of a turbulent channel flow. The proposed multi-layer perceptrons were fed with information locally related to the actuation point. In one of the studied cases, the flow was completely relaminarised, demonstrating another class of flows that can be controlled using a machine learning framework. Another way to leverage NNs for flow control was proposed by \cite{morton2018deep}, who modelled the dynamics of a cylinder flow with models trained to learn state mappings to span a Koopman invariant subspace. Significant vortex shedding attenuation was achieved by feeding flow images into a convolutional neural network (CNN) model iterated through time within a finite horizon, which was used for model predictive control (MPC). Other techniques employed for flow control include network analysis for control design \citep{yeh2021network}, sparse identification of nonlinear dynamics (SINDy) with control \citep{brunton2016sparse}, real-time extremum seeking \citep{deda2021extremum,payne2024co} and direct opposition control \citep{choi1994active}.

While these previous studies presented innovative control design capable of accomplishing important goals, there are barriers that hinder their implementation in real-world systems. One of the most important limitations is the sensor setup. In experiments, online sensing is usually limited to a few pressure and skin friction probes that need to be located along the walls, since velocity measurements in the flow field, e.g., through particle image velocimetry (PIV), can be both expensive and computationally prohibitive in real time. While velocity measurements with hotwire probes can be feasible for online sensing, these also only typically allow for a smaller number of sensors, and can be disruptive to the downstream flow field. For these reasons, open-loop control strategies are also often explored to manipulate fluid flows. 
In such cases, the optimisation of flow variables is done by tuning the actuator to work offline, regardless of the real-time flow signals. For example, through large-eddy simulations, \cite{visbal2018control}, \cite{Brener} and \cite{Feitosa2025} studied the open-loop control of airfoils under dynamic stall. By setting specific actuation frequencies for oscillatory jets near the leading edge, significant reductions were observed in the phase-averaged drag. 
Genetic algorithms (GAs) have also been applied in experimental studies to find the best actuation setups to optimize flow variables. \cite{zigunov2022bluff} employed a GA to find a subset of actuators from a set of candidates to reduce drag in a flow over a bluff body, while \cite{zigunov2023multiaxis} used a similar approach to  maximise the thrust vectoring angle in a supersonic jet. \cite{raibaudo2020machine} used GAs to control the flow past a triangular array of cylinders known as a fluidic pinball, targeting either drag reduction or flow symmetry by setting the rotation speed of the cylinders.

Another way to approach sensor limitations is by discovering optimal sensor placement to reduce the required number of probes. For example, \cite{paris2021robust} introduced a reinforcement learning approach to develop drag reduction control strategies while optimizing sensor placement using an L0 regularization scheme. \cite{deda2024neural} trained NN models for control design using an L1 regularization approach to reduce the number of sensors used to evolve the system states in time. Flow reconstruction from limited sensor data is also an area of interest within fluid dynamics. Different techniques have been used to approach this problem, such as gappy proper orthogonal decomposition \citep{willcox2006unsteady}, a pivoted QR decomposition of an identified set of basis functions \citep{manohar2018data}, data-driven dynamical models with Kalman filters \citep{sashittal2021data,graff2023information}, and decoder neural networks \citep{williams2022data}. Turbulent channel flow reconstruction from wall probes that read pressure and shear stresses was performed by \cite{nakamura2022robust}. Resolvent analysis is also a powerful tool for flow reconstruction. Its applications include the inference of statistical properties and reconstruction of flow fields, allowing for causal approximations --- either through truncation of optimal kernels or by optimal modelling through a Wiener-Hopf procedure --- of the involved transfer functions \citep{martini2020resolvent,towne2018,amaral2021resolvent}. 

Reconstruction problems such as estimating velocity fields from particle images and super-resolution have also been investigated \citep{morimoto2022generalization, miotto2025}.
The latter approach involves techniques developed for upscaling low-resolution flow images \citep{fukami2023super}. The super-resolution analysis of the flow past two cylinders distanced by a varying gap was done by \cite{morimoto2022generalization}, where low-resolution images are upscaled using CNNs to provide a flow field with finer details. \cite{fukami2019} implemented an NN architecture that combines a CNN upscaler with downsampled skip connections and a CNN with multi-scale filters to infer details of a turbulent flow. \cite{page2024super} proposed a super-resolution estimation approach where NNs are trained using backpropagation through time (BPTT) without the need for high resolution reference data by leveraging a differentiable dynamic flow model. Starting from a coarse initial condition, their approach stores memory from past estimations and produces accurate flow estimations after a few iterations over time. 

In the present work, we approach flow estimation from limited sensor data by leveraging topologies commonly used in control theory. For example, two well established state estimation approaches integral to modern control theory are the Luenberger observer and the Kalman filter. Taking inspiration from such approaches, we propose a dynamic estimation of the flow variables --- as opposed to a static reconstruction --- by leveraging models with memory signals that correspond to the plant state space. The implementation of closed-loop NN-based state observers, or simply NN observers (NNOs), can be done through approaches analogous to neural network controllers (NNCs). \cite{deda2024neural}, for example, proposed the training of NNCs via BPTT, where the controller is trained to stabilize a neural network surrogate model (NNSM) by approaching an equilibrium point within a finite time horizon. A similar framework was presented by \cite{yadaiah2006neural}, who leveraged BPTT to train state observers for simple plants. By testing their technique with low-dimensional nonlinear systems, the authors showed that the NNs were able to outperform implementations of the extended Kalman filter. 

Here, we propose the implementation of NNOs to enable dynamic output feedback control of flows by dynamically estimating states from limited sensor data. To do so, we leverage previously trained NNSMs, which contain states whose sensing is not feasible in real world systems. Instead of assuming real-time feedback of the states, we leverage real-time data from more realistic sensors to infer the states required to feed the controllers. The present methodology is inspired by the Luenberger observer, but machine learning tools are used to replace the systems from the traditional linear approach, which can present prohibitive limitations when working with strongly nonlinear systems such as fluid flows. Other approaches for the generalization of the Luenberger observer include the application of NNs for learning mappings to provide a nonlinear observer considering continuous-time input-affine systems \citep{ramos2020numerical}. Furthermore, \cite{zeitz1987extended} and \cite{afri2016state} reported extensions of the Luenberger observer applied to low-dimensional single-input single-output systems. In the realm of fluid flows, NNs can be used to provide more general approaches that are not necessarily limited to very strict assumptions. This comes, at the same time, with the cost of not providing formal convergence proofs, but the complex dynamics of unstable flows make it necessary to explore black-box techniques. Current literature provides estimation techniques that employ flow solvers or surrogate models to complete information by embedding dynamics \citep{page2024super, zhong2023sparse}.




The proposed framework is tested with three different nonlinear systems, namely a modified Kuramoto-Sivashinsky equation, a confined cylinder flow previously studied by \cite{rabault2019artificial}, and experimental data obtained from a turbulent boundary layer. The first two cases are tested within numerical simulations, and the observers are employed to estimate the states that are in turn fed to a pre-trained stabilizing controller. Real-time measurements for state estimation are obtained assuming adverse conditions, such as insufficient sampling time to match the discrete plant model, reduced number of sensors, and different types of random noise. For the boundary layer case, we use the experimental data provided by \cite{zigunov2020detailed} to train all required NNs. For this setup, specialized NN architectures are proposed to process PIV data. Low resolution images of the flow velocities are employed for the estimation of flow variables, in an approach that differs from current super-resolution techniques by leveraging NNSM predictions. Since the available turbulent boundary layer dataset is from a previous experiment to which our group does not have access, the tests are restricted to flow state estimation without control. 
The remaining sections of the present work are organised as follows: \S\ref{sec:methodology} discusses the methodology employed for flow state estimation through limited sensor data; \S\ref{sec:studycases} presents the plant setups studied; \S\ref{sec:results} reports the results obtained for each case; and \S\ref{sec:conclusions} discusses the main achievements and limitations of the proposed methodology. In the scope of the present work, the terms ``observer'' and ``estimator'' are used interchangeably.

\section{Methodology} \label{sec:methodology}

In this section, the mathematical tools utilised in the implementation of the NNOs are presented. First, a review on the discrete Luenberger observers is provided to introduce the main ideas that inspired the choices regarding the neural network approach. A nonlinear scheme analogous to the Luenberger observers is then presented. Finally, a modified structure for the estimator loop is introduced, in order to make the implementation feasible in the context of the studied systems, which are described in \S\ref{sec:studycases} The training setup for the NNOs is then presented. For a general overview on state observers, we refer the reader to \cite{luenberger2007observing} and \cite{friedland2012control} for a review on observers.

\subsection{Discrete Luenberger observer}

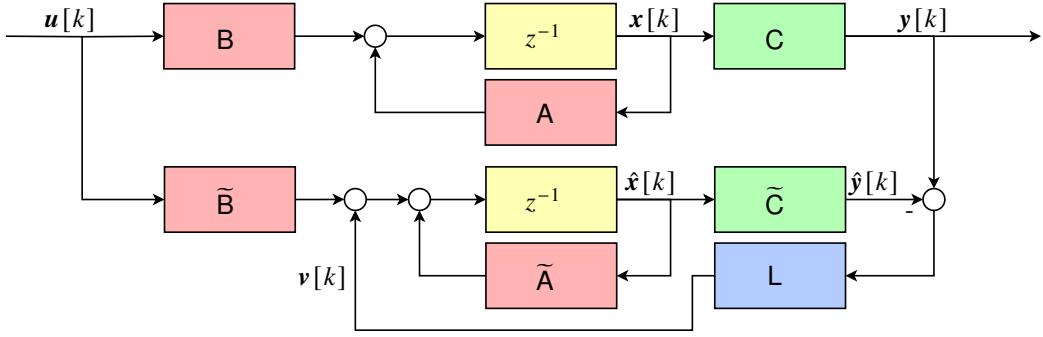
\begin{figure}
\centering
\begin{overpic}[width=0.9\linewidth]{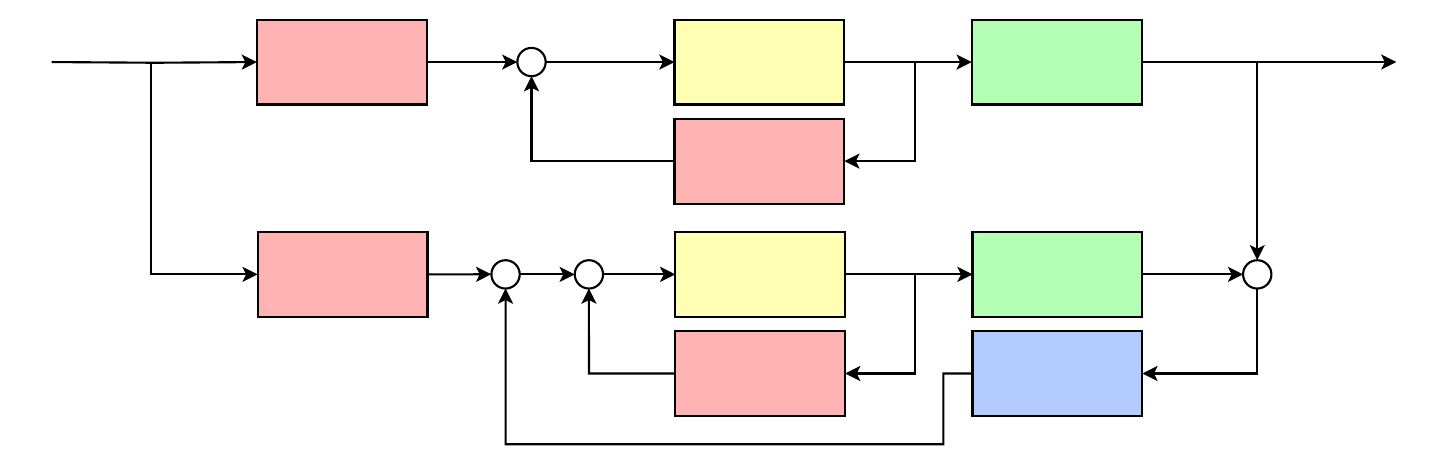}
\put(7,29.1){\scalebox{1.0}{$\bm{u}[k]$}} 
\put(59.4,29.1){\scalebox{1.0}{$\bm{x}[k]$}} 
\put(83.4,29.1){\scalebox{1.0}{$\bm{y}[k]$}} 
\put(59,14.5){\scalebox{1.0}{$\hat{\bm{x}}[k]$}} 
\put(79,14.5){\scalebox{1.0}{$\hat{\bm{y}}[k]$}} 
\put(84,12){-} 
\put(29.7,6){\scalebox{1.0}{$\bm{v}[k]$}} 
\put(50,27.4){\scalebox{1.0}{$z^{-1}$}} 
\put(50,12.7){\scalebox{1.0}{$z^{-1}$}} 
\put(51,20.5){\scalebox{1.0}{$\mathsfbi{A}$}} 
\put(51,5.6){\scalebox{1.0}{$\widetilde{\mathsfbi{A}}$}} 
\put(71.7,6){\scalebox{1.0}{$\mathsfbi{L}$}} 
\put(71.5,27.2){\scalebox{1.0}{$\mathsfbi{C}$}} 
\put(71.5,12.4){\scalebox{1.0}{$\widetilde{\mathsfbi{C}}$}} 
\put(22.4,27.2){\scalebox{1.0}{$\mathsfbi{B}$}} 
\put(22.4,12.4){\scalebox{1.0}{$\widetilde{\mathsfbi{B}}$}} 
\end{overpic}
\caption{Block diagram representing 
the discrete Luenberger observer for a linear dynamical system. The plant model is simulated in real time as the real system evolves. The correction signal $\bm{v}[k]$ is produced to rectify errors due to initial condition, plant imperfections and unmodelled disturbances.}
\label{fig:loop_linear}
\end{figure}

Consider the discrete-time invariant linear dynamical system described by
\begin{align}
    \bm{x}[k+1] &= \mathsfbi{A} \bm{x}[k] + \mathsfbi{B} \bm{u}[k] \mbox{ ,}\\
    \bm{y}[k] &= \mathsfbi{C} \bm{x}[k] \mbox{ ,}
\end{align}
where $\bm{x}[k]$ is the state vector at discrete time $k$, $\bm{u}[k]$ is the control input vector, and $\bm{y}[k]$ is the measurable output vector. Matrices $\mathsfbi{A}$, $\mathsfbi{B}$, and $\mathsfbi{C}$ are the state, input-to-state, and state-to-output matrices, respectively. Given that $\bm{u}[k]$ is known, a discrete Luenberger observer can be employed to estimate the plant states by solving the corresponding system of difference equations
\begin{align}
    \hat{\bm{x}}[k+1] &= \widetilde{\mathsfbi{A}} \hat{\bm{x}}[k] + \widetilde{\mathsfbi{B}} \bm{u}[k] + \bm{v}[k] \mbox{ ,}\\
    \bm{v}[k] &= \mathsfbi{L}(\bm{y}[k] - \hat{\bm{y}}[k]) \mbox{ ,}\\
    \hat{\bm{y}}[k] &= \widetilde{\mathsfbi{C}}\hat{\bm{x}}[k] \mbox{ ,}
\end{align}
where $\mathsfbi{L}$ is the estimator gain and $\widetilde{\mathsfbi{A}}$, $\widetilde{\mathsfbi{B}}$, and $\widetilde{\mathsfbi{C}}$ are approximations of the real plant parameter matrices $\mathsfbi{A}$, $\mathsfbi{B}$, and $\mathsfbi{C}$. The estimated states $\hat{\bm{x}}[k+1]$ are predicted from the current estimate $\hat{\bm{x}}[k]$ and from the known control input $\bm{u}[k]$. A closed-loop correction law uses the sensed variables $\bm{y}[k]$ to correct $\hat{\bm{x}}[k+1]$ based on the estimated output $\hat{\bm{y}}[k] = \widetilde{\mathsfbi{C}} \hat{\bm{x}}[k]$. The block diagram for this classic setup is presented in figure~\ref{fig:loop_linear}, where $z^{-1}$ represents one time step delay with a notation that comes from the Z-transform frequency domain. If the system is observable and $\mathsfbi{L}$ is designed (e.g., through pole placement) such that $\bm{x} - \hat{\bm{x}} \to 0$ as $k \to \infty$, a perfect model (i.e., $\widetilde{\mathsfbi{A}} = \mathsfbi{A}$, $\widetilde{\mathsfbi{B}} = \mathsfbi{B}$, and $\widetilde{\mathsfbi{C}} = \mathsfbi{C}$) would tendentially bring $\bm{v}[k]$ to zero, and correction would no longer be needed.

Another famous approach to estimators is the Kalman filter, which consists of an algorithm similar to the Luenberger observer. The main difference is that the estimator gain (here represented by $\mathsfbi{L}$) is updated every time step according to the covariances related to process and measurement noises. The Kalman gain represents the optimal correction factor, assuming noise sources are Gaussian and white. Although the focus of the present work is to estimate states from limited sensor data without noise, we also present possible modifications to the NNO training to encompass measurement noise and insuficient sampling rate.

\subsection{A nonlinear generalisation of the Luenberger observer}



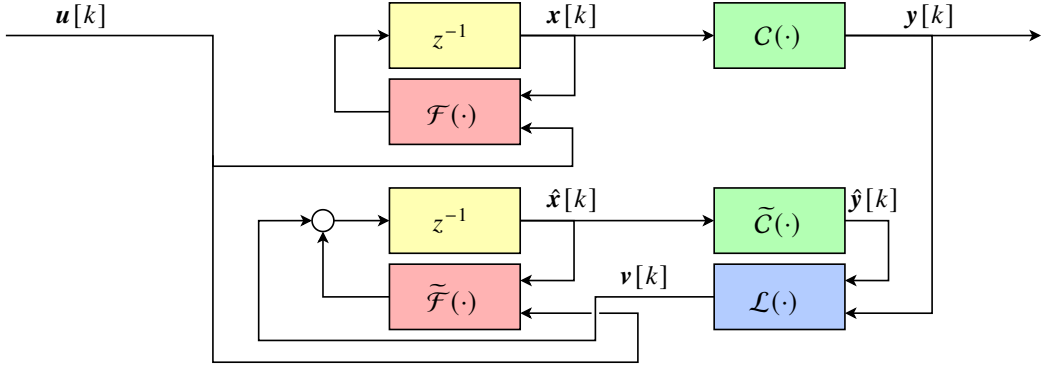
\begin{figure}
\centering
\begin{overpic}[width=0.9\linewidth]{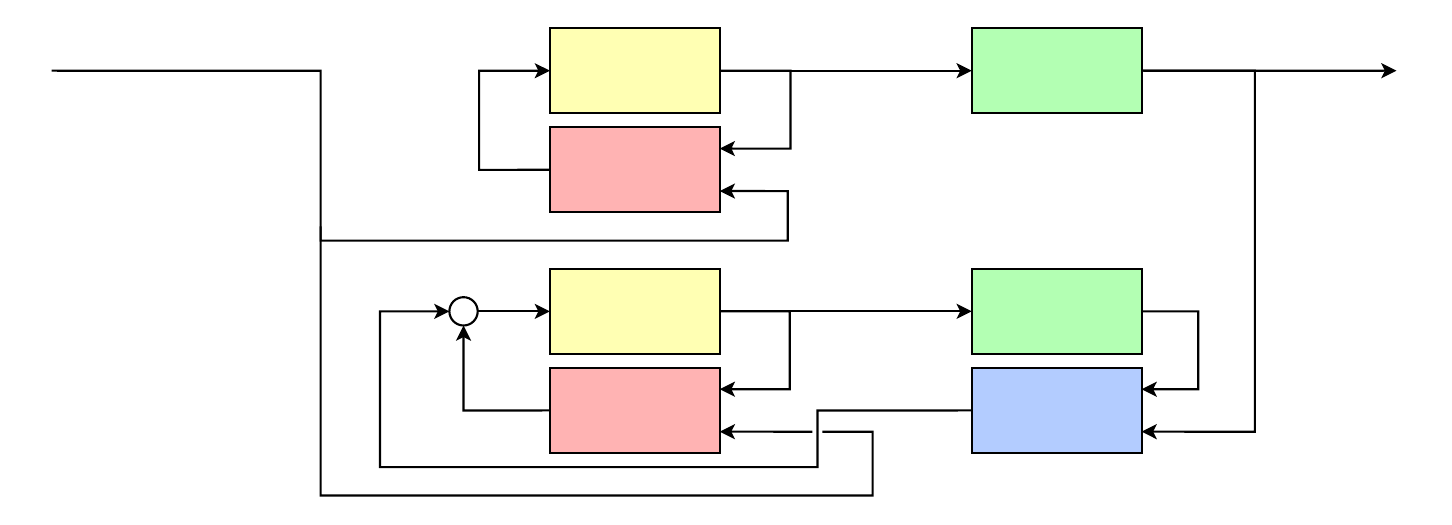}
\put(8,32.4){\scalebox{1.0}{$\bm{u}[k]$}} 
\put(52,32.4){\scalebox{1.0}{$\bm{x}[k]$}} 
\put(84,32.4){\scalebox{1.0}{$\bm{y}[k]$}} 
\put(52,15.9){\scalebox{1.0}{$\hat{\bm{x}}[k]$}} 
\put(79,15.9){\scalebox{1.0}{$\hat{\bm{y}}[k]$}} 
\put(58.5,9){\scalebox{1.0}{$\bm{v}[k]$}} 
\put(41.7,30.3){\scalebox{1.0}{$z^{-1}$}} 
\put(41.7,13.8){\scalebox{1.0}{$z^{-1}$}} 
\put(41.2,23.5){\scalebox{1.0}{$\mathcal{F}(\cdot)$}} 
\put(41.2,6.7){\scalebox{1.0}{$\widetilde{\mathcal{F}}(\cdot)$}} 
\put(70,6.7){\scalebox{1.0}{$\mathcal{L}(\cdot)$}} 
\put(70.5,30.5){\scalebox{1.0}{$\mathcal{C}(\cdot)$}} 
\put(70.5,13.7){\scalebox{1.0}{$\widetilde{\mathcal{C}}(\cdot)$}} 
\end{overpic}
\caption{Block diagram representing a nonlinear generalisation of the discrete Luenberger observer. The colours are chosen to highlight the analogy with the classic approach for linear systems, depicted in figure~\ref{fig:loop_linear}.}
\label{fig:loop_nonlinear}
\end{figure}

Consider the nonlinear dynamical system described by
\begin{align}
    \bm{x}[k+1] &= \mathcal{F}(\bm{x}[k], \bm{u}[k]) \mbox{ ,} \\
    \bm{y}[k] &= \mathcal{C}(\bm{x}[k]) \mbox{ ,}
\end{align}
where $\mathcal{F}(\bm{x}, \bm{u})$ is a general nonlinear function that governs the dynamics of the states, and $\mathcal{C}(\bm{x})$ is a general nonlinear function that maps the states to the output variable space. If approximations $\widetilde{\mathcal{F}}(\bm{x}, \bm{u}) \approx \mathcal{F}(\bm{x}, \bm{u})$ and $\widetilde{\mathcal{C}}(\bm{x}, \bm{u}) \approx \mathcal{C}(\bm{x}, \bm{u})$ are available, an estimator loop can be built through the implementation of the difference equations
\begin{align}
    \hat{\bm{x}}[k+1] &= \widetilde{\mathcal{F}}(\hat{\bm{x}}[k], \bm{u}[k]) + \bm{v}[k] \mbox{ ,} \\
    \bm{v}[k] &= \mathcal{L}(\bm{y}[k], \hat{\bm{y}}[k]) \mbox{ ,} \\
    \hat{\bm{y}}[k] &=  \widetilde{\mathcal{C}}(\hat{\bm{x}}[k])\mbox{ .} 
\end{align}
The structure is similar to the Luenberger approach, which can be verified by comparing the equations, or by noticing the similarities between the block diagrams in figures~\ref{fig:loop_linear} and \ref{fig:loop_nonlinear}. In both cases, the convergence of $\hat{\bm{x}}$ to $\bm{x}$ would also imply the convergence of $\hat{\bm{y}}$ to $\bm{y}$ and, since the trajectories for the estimated and actual states are constrained by the same dynamics (assuming the plant model is perfect), $\bm{v}[k]$ approximates zero and the estimation would be based only on predictions. In the present work, the nonlinear operators $\widetilde{\mathcal{F}}$, $\widetilde{\mathcal{C}}$, and $\mathcal{L}$ are implemented as neural networks. Notice that the process of computing $\bm{u}[k]$ is omitted, but it could be represented as 
\begin{align}
    \bm{u}[k] &= \bm{u}_\mathrm{c}[k] + \bm{u}_\mathrm{o}[k] \mbox{ ,}\\
    \bm{u}_\mathrm{c}[k] &= \mathcal{K}(\hat{\bm{x}}) \mbox{ ,}
\end{align}
where $\bm{u}_\mathrm{c}[k]$ and $\bm{u}_\mathrm{o}[k]$ are the closed-loop and open-loop components of the control input, respectively. The nonlinear function $\mathcal{K}$ is implemented as an NNC, and is not represented in figures~\ref{fig:loop_linear} and \ref{fig:loop_nonlinear} for simplicity. 

\subsection{Training setup}

The proposed observer loop consists of three main neural networks: the NNSM $\widetilde{\mathcal{F}}$, the output model $\widetilde{\mathcal{C}}$, and the NNO $\mathcal{L}$. For closed-loop control cases, the loop will also contain the NNC $\mathcal{K}$. To train the NNSM and the NNC, the methodology proposed by \cite{deda2024neural} is followed. Thus, these models have already been trained, and our objective is to train
$\widetilde{\mathcal{C}}$ and $\mathcal{L}$. All neural networks involved in this work have a scaling layer to normalize their inputs such that the average and standard deviation of the data entering the first hidden layer are zero and one, respectively.

\subsubsection{Output Model}

To train $\widetilde{\mathcal{C}}$, a simple supervised learning approach is followed. First, let us consider numerical simulations, where all time-resolved system variables are available. In this case, one can simply collect $\bm{u}$, $\bm{x}$, and $\bm{y}$ and find the best fit through backpropagation. The loss function chosen is
\begin{equation}
    l_\mathrm{\widetilde{\mathcal{C}}} = \frac{1}{n_d} \sum_{i=1}^{n_d} \norm{ \bm{y}_i - \widetilde{\mathcal{C}}(\bm{x}_i)} _2^2 + \lambda_{\widetilde{\mathcal{C}}} \norm{\bm{w}_{\widetilde{\mathcal{C}}}}_2^2 \mbox{ ,}
\end{equation}
where $\bm{w}_{\widetilde{\mathcal{C}}}$ are the weights of the output model, $\lambda_{\widetilde{\mathcal{C}}} $ tunes the strength of the L2 regularization, and $n_d$ is the number of samples in the input/output training batch.

For experimental applications, the training data for $\widetilde{\mathcal{C}}$ must be obtained through alternative means, as full-state information is typically not available. One possible approach is to use models derived from equivalent numerical simulations to approximate $\mathcal{C}$. Alternatively, specialized sensors can be used to directly measure the system states during the training phase. These sensors are solely required for dataset collection and are not necessary for the observer loop after training. For instance, in the case of estimating velocity fields (states) from wall pressure measurements (outputs), particle image velocimetry could be used to construct the training dataset by capturing the velocity field while simultaneously recording the wall pressure. In this case, the data required for learning the static map $\widetilde{\mathcal{C}}$ does not need to be time-resolved at strict sample rates. Once trained, the network should be able to estimate velocity fields based only on wall pressure measurements, eliminating the need for PIV in real-time operation.

\subsubsection{Neural Network Observer}

\begin{figure}
\centering
\begin{overpic}[width=0.8\linewidth]{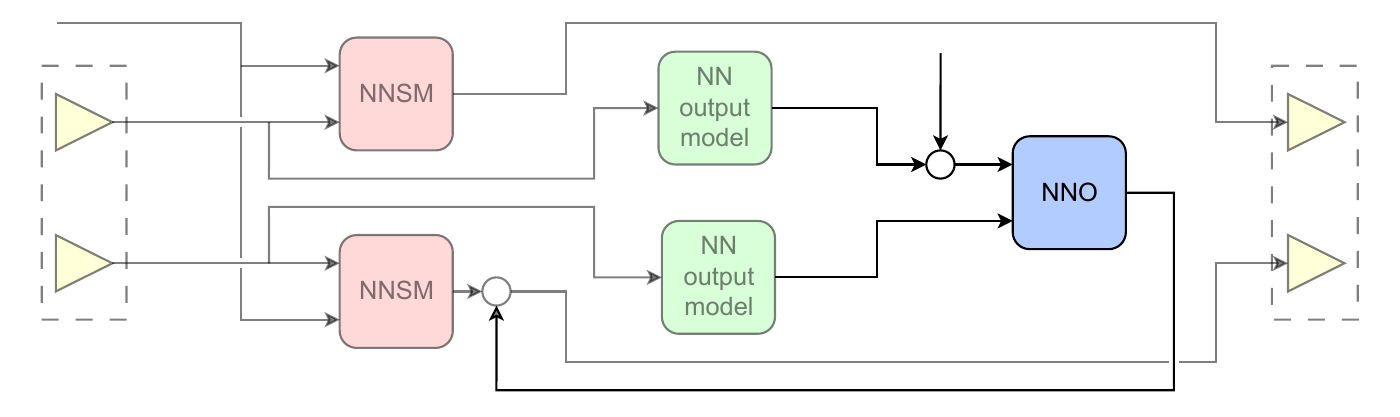}
\put(8,30){\scalebox{0.85}{\transparent{0.5} $\bm{u}[k]$}}
\put(9.5,22){\scalebox{0.85}{\transparent{0.5} $\bm{x}[k]$}}
\put(9.5,12){\scalebox{0.85}{\transparent{0.5} $\hat{\bm{x}}[k]$}}
\put(90,26){\scalebox{0.85}{\transparent{0.5} $\bm{x}[k+1]$}}
\put(90,4.3){\scalebox{0.85}{\transparent{0.5} $\hat{\bm{x}}[k+1]$}}
\put(56,23){\scalebox{0.85}{$\bm{y}[k]$}}
\put(68,23){\scalebox{0.85}{$\bm{y}_n[k]$}}
\put(56,7.5){\scalebox{0.85}{$\hat{\bm{y}}[k]$}}
\put(81,17){\scalebox{0.85}{$\bm{v}[k]$}}
\put(0,5){\rotatebox{90}{\scalebox{0.74}{From previous time step}}}
\put(98,8.5){\rotatebox{90}{\scalebox{0.74}{To next time step}}}
\end{overpic}
\caption{Schematic of a single iteration of the observer training loop. The NNSM and output model weights are frozen and only the NNO ones are updated. The hat notation ($\hat{x}$ and $\hat{y}$) indicates estimated signals.}
\label{fig:nn_step}
\end{figure}
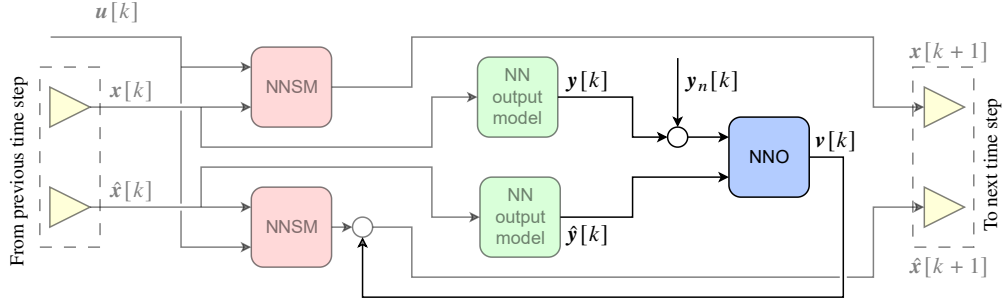

Now that $\widetilde{\mathcal{C}}$ is trained, a closed-loop approach is proposed for training $\mathcal{L}$. The idea is to use a recurrent auxiliary configuration for training the NNO, in line with the method proposed by \cite{yadaiah2006neural}. This approach is analogous to the finite-horizon training strategies commonly found in closed-loop control literature \citep{bieker2020deep,deda2023backpropagation}. Figure~\ref{fig:nn_step} illustrates the structure of a single iteration of the observer loop, where $y_n$ is a measurement noise source, further detailed in this section. The colour scheme is consistent with that of figure~\ref{fig:loop_nonlinear}, with analogous blocks highlighted similarly. 
The NNSM and the output model are each used twice during training: once to represent the real system and once to represent the observer predictor. Their weights are kept fixed and only the weights of the NNO are updated. After training, the upper branch (representing the real system) is replaced by the actual plant during testing.

\begin{figure}
\centering
\begin{overpic}[width=0.98\linewidth]{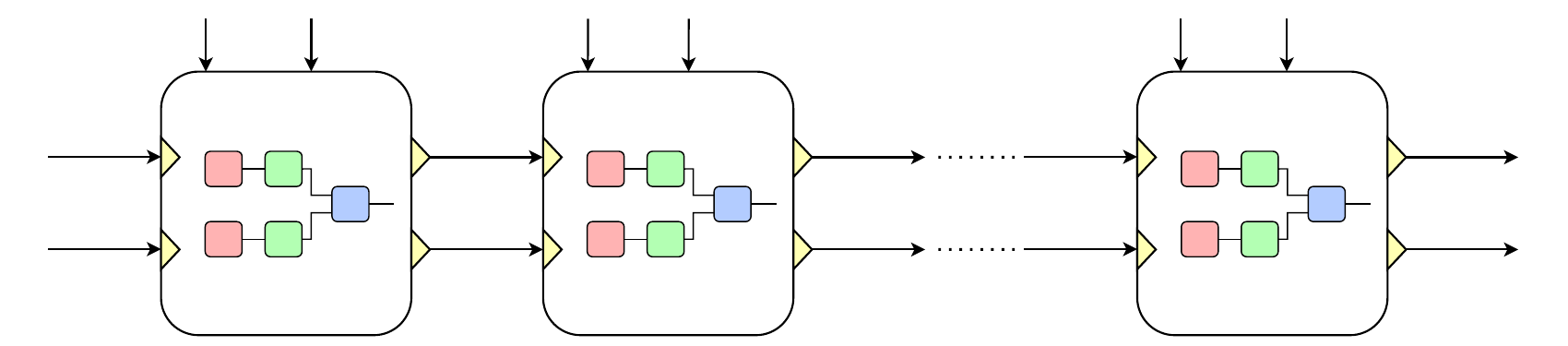}
\put(7.8,19){\scalebox{0.85}{$\bm{u}_{\mathrm{o}}[0]$}}
\put(32.2,19){\scalebox{0.85}{$\bm{u}_{\mathrm{o}}[1]$}}
\put(66.5,19){\scalebox{0.85}{$\bm{u}_{\mathrm{o}}[n_h{-}1]$}}
\put(20.8,19){\scalebox{0.85}{$\bm{y}_n[0]$}}
\put(45.0,19){\scalebox{0.85}{$\bm{y}_n[1]$}}
\put(83.1,19){\scalebox{0.85}{$\bm{y}_n[n_h-1]$}}
\put(4,13.0){\scalebox{0.85}{$\bm{x}[0]$}}
\put(4,7.0){\scalebox{0.85}{$\hat{\bm{x}}[0]$}}
\put(28.4,13.0){\scalebox{0.85}{$\bm{x}[1]$}}
\put(28.4,7.0){\scalebox{0.85}{$\hat{\bm{x}}[1]$}}
\put(52.8,13.0){\scalebox{0.85}{$\bm{x}[2]$}}
\put(52.8,7.0){\scalebox{0.85}{$\hat{\bm{x}}[2]$}}
\put(63.2,13.0){\scalebox{0.85}{$\bm{x}[n_h{-}1]$}}
\put(63.2,7.0){\scalebox{0.85}{$\hat{\bm{x}}[n_h{-}1]$}}
\put(90.6,13.0){\scalebox{0.85}{$\bm{x}[n_h]$}}
\put(90.6,7.0){\scalebox{0.85}{$\hat{\bm{x}}[n_h]$}}
\end{overpic}
\caption{Unrolled observer loop over a finite horizon of length $n_h$, with each neural network block representing the complete NNO loop.}
\label{fig:nn_rec}
\end{figure}

The single-step structure is unrolled along a finite horizon, as illustrated in figure~\ref{fig:nn_rec}, where $n_h$ denotes the horizon length. The training dataset includes initial conditions $\bm{x}[0]$, $\bm{v}[0]$, and $\hat{\bm{x}}[0]$, as well as open-loop input sequences of length $n_h$: $\bm{u}_\mathrm{o}[0], \bm{u}_\mathrm{o}[1], \dots, \bm{u}_\mathrm{o}[n_h{-}1]$. A measurement noise source $y_n[k]$ can be added to train the observer loop under conditions more reminiscent of those expected in real applications. If the measurement noise for a given application can be modelled (e.g., white or time-correlated Gaussian noise), that same type of noise can by employed during training. In the present work, two types of noise sources are employed: Gaussian white with zero mean and standard deviation $\sigma$; and time-correlated noise obtained by filtering a white Gaussian noise source $y_n^*$ such that
\begin{equation}
\label{eq:corrnoise}
{y_n[k] = y_n^*[k] + \beta y_n[k-1]} \mbox{ ,}
\end{equation}
where $0 \leq \beta \leq 1 $ determines the level of correlation in time.

\begin{figure}
\centering
\begin{overpic}[width=0.98\linewidth]{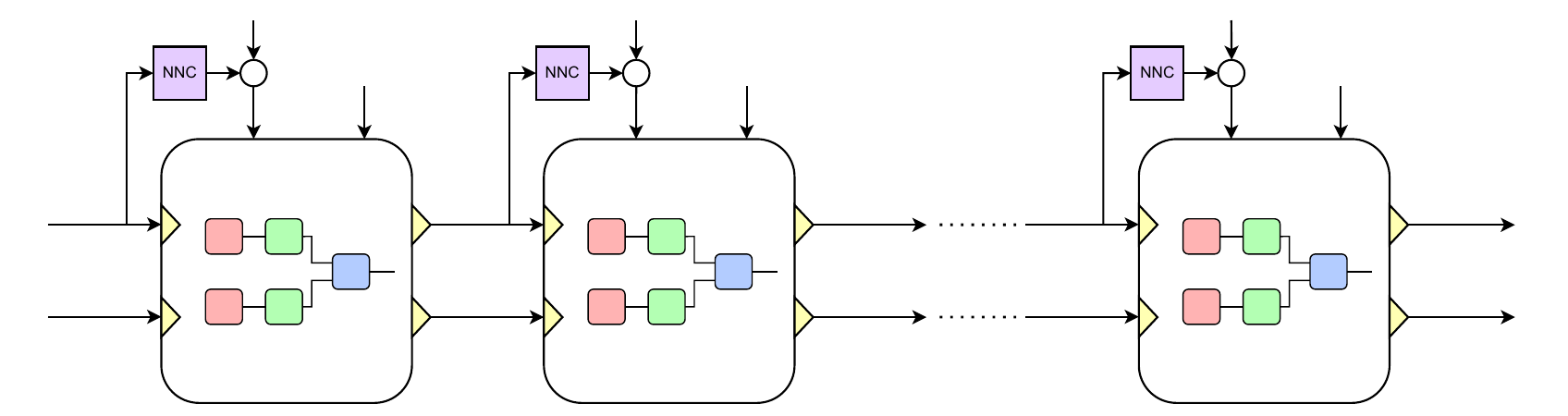}
\put(16.8,24){\scalebox{0.85}{$\bm{u}_{\mathrm{o}}[0]$}}
\put(41.2,24){\scalebox{0.85}{$\bm{u}_{\mathrm{o}}[1]$}}
\put(79.0,24){\scalebox{0.85}{$\bm{u}_{\mathrm{o}}[n_h{-}1]$}}
\put(24,19){\scalebox{0.85}{$\bm{y}_n[0]$}}
\put(48.2,19){\scalebox{0.85}{$\bm{y}_n[1]$}}
\put(86.3,19){\scalebox{0.85}{$\bm{y}_n[n_h-1]$}}
\put(4,13.0){\scalebox{0.85}{$\bm{x}[0]$}}
\put(4,7.0){\scalebox{0.85}{$\hat{\bm{x}}[0]$}}
\put(28.4,13.0){\scalebox{0.85}{$\bm{x}[1]$}}
\put(28.4,7.0){\scalebox{0.85}{$\hat{\bm{x}}[1]$}}
\put(52.8,13.0){\scalebox{0.85}{$\bm{x}[2]$}}
\put(52.8,7.0){\scalebox{0.85}{$\hat{\bm{x}}[2]$}}
\put(90.6,13.0){\scalebox{0.85}{$\bm{x}[n_h]$}}
\put(90.6,7.0){\scalebox{0.85}{$\hat{\bm{x}}[n_h]$}}
\end{overpic}
\caption{Finite-horizon observer training loop with controller (NNC) inclusion.}
\label{fig:nn_reccon}
\end{figure}
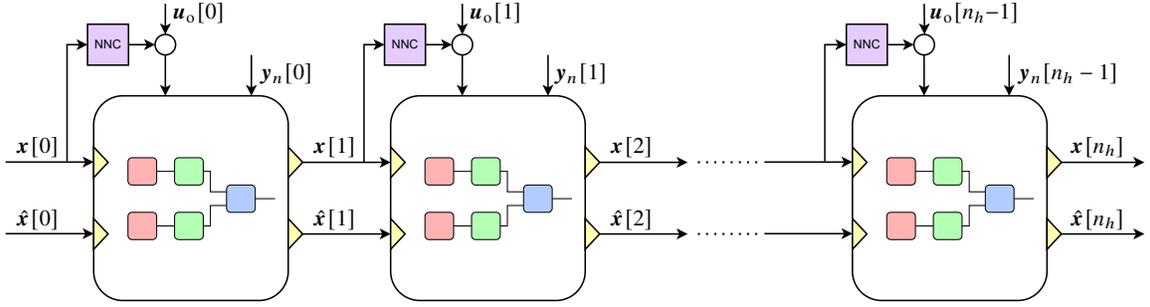

The loss function used to train the NNO is the element-wise average of the expression
\begin{align}
l_{\mathcal{L}} &= \frac{1}{n_d n_h} \sum_{i=1}^{n_d} \sum_{k=1}^{n_h} \left( \norm{ \bm{e}_{y,i}[k] }_2^2 + \alpha_x \norm{ \bm{e}_{x,i}[k] }_2^2 + \alpha_v \norm{ \bm{v}_i[k] }_2^2 \right) + \lambda_{\mathcal{L}} \norm{\bm{w}_{\mathcal{L}}}_2^2 \mbox{ ,} \\
\bm{e}_y[k] &= \bm{y}[k] - \hat{\bm{y}}[k] \mbox{ ,} \\
\bm{e}_x[k] &= \bm{x}[k] - \hat{\bm{x}}[k] \mbox{ ,}
\end{align}
where $\alpha_x$ and $\alpha_v$ are coefficients that control the penalization of state errors and correction signals, respectively, and $\lambda_{\mathcal{L}}$ controls the amount of L2 regularization on the NNO weights $\bm{w}_{\mathcal{L}}$. Index $i$ refer to the signals propagated from the $i$-th initial condition in the dataset. For training, the initial states $\bm{x}[0]$ are sampled from the true system (either numerical or experimental), and $\hat{\bm{x}}[0]$ uses the same values but shuffled to prevent the observer from always encountering the trivial solution $\bm{v}[k]=\bm{0}$. The control signals used during training are random excerpts from those employed to generate the NNSM training data . Furthermore, figure~\ref{fig:nn_reccon} shows the inclusion of the NNC within the loop to ensure the observer is trained in state-space regions relevant to closed-loop operation. In the present work, two different systems with control inputs are studied, namely a modified KS equation and a confined cylinder flow with jet actuation. In these cases, the training loop includes the NNC as shown in figure \ref{fig:nn_reccon}. On the other hand, for the third system studied, a turbulent boundary layer that does not involve closed-loop control, only open-loop signals (obtained from the dataset) are involved as depicted in figure \ref{fig:nn_rec}.

Training the NNO can be challenging due to the potential for instability when operating in closed loop. If $n_h$ is set too large, the feedback dynamics --- especially with untrained, randomly initialized weights --- may become unstable and result in exponential divergence. In practice, values such as $n_h = 15$ were found to be sufficient to cause training difficulties in some cases. To mitigate this, the horizon length $n_h$ is gradually increased over training: beginning with a small $n_h$ and incrementing it every few epochs. Both the initial $n_h$ and the increment schedule are treated as hyperparameters.

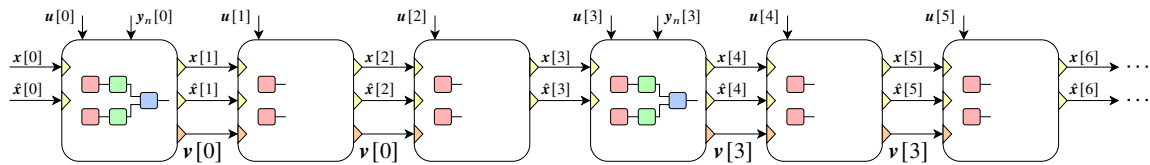
\begin{figure}
\centering
\begin{overpic}[width=0.98\linewidth]{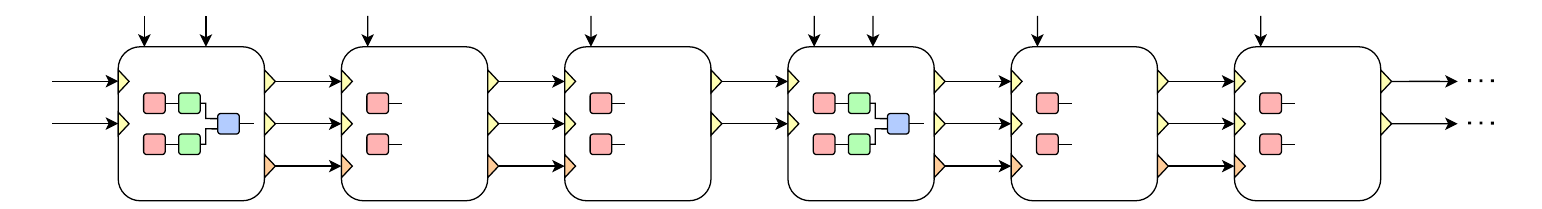}
\put(17.6,0.8){\scalebox{0.85}{$\bm{v}[0]$}}
\put(32.1,0.8){\scalebox{0.85}{$\bm{v}[0]$}}
\put(61.2,0.8){\scalebox{0.85}{$\bm{v}[3]$}}
\put(75.6,0.8){\scalebox{0.85}{$\bm{v}[3]$}}
\put(3.7,8.7){\scalebox{0.60}{$\bm{x}[0]$}}
\put(18.1667,8.7){\scalebox{0.60}{$\bm{x}[1]$}}
\put(32.6333,8.7){\scalebox{0.60}{$\bm{x}[2]$}}
\put(47.1000,8.7){\scalebox{0.60}{$\bm{x}[3]$}}
\put(61.5667,8.7){\scalebox{0.60}{$\bm{x}[4]$}}
\put(76.0333,8.7){\scalebox{0.60}{$\bm{x}[5]$}}
\put(90.5,8.7){\scalebox{0.60}{$\bm{x}[6]$}}
\put(3.7,6){\scalebox{0.60}{$\hat{\bm{x}}[0]$}}
\put(18.1667,6){\scalebox{0.60}{$\hat{\bm{x}}[1]$}}
\put(32.6333,6){\scalebox{0.60}{$\hat{\bm{x}}[2]$}}
\put(47.1000,6){\scalebox{0.60}{$\hat{\bm{x}}[3]$}}
\put(61.5667,6){\scalebox{0.60}{$\hat{\bm{x}}[4]$}}
\put(76.0333,6){\scalebox{0.60}{$\hat{\bm{x}}[5]$}}
\put(90.5,6){\scalebox{0.60}{$\hat{\bm{x}}[6]$}}
\put(6.3,12){\scalebox{0.60}{$\bm{u}[0]$}}
\put(20.7667,12){\scalebox{0.60}{$\bm{u}[1]$}}
\put(35.2333,12){\scalebox{0.60}{$\bm{u}[2]$}}
\put(49.7000,12){\scalebox{0.60}{$\bm{u}[3]$}}
\put(64.1667,12){\scalebox{0.60}{$\bm{u}[4]$}}
\put(78.6333,12){\scalebox{0.60}{$\bm{u}[5]$}}
\put(13.8,12){\scalebox{0.60}{$\bm{y}_n[0]$}}
\put(57.2,12){\scalebox{0.60}{$\bm{y}_n[3]$}}
\end{overpic}
\caption{Training structure with time skips and $\Delta k = 3$. Signal $\bm{v}[k]$ is propagated for time steps where measurements are absent.}
\label{fig:nn_recskip}
\end{figure}

Another advantage of neural networks as observers is the possibility of considering sensors with sampling rates lower than those required for control. By choosing an integer $\Delta k > 1$, the training structure presented in figure \ref{fig:nn_recskip} can be built. In this approach, $\bm{v}[k]$ is only computed every $\Delta k$ steps, and propagated to all iterations until the next computation. Therefore, the lower sensor sampling time is taken into account during training. More information regarding the legacy NNSM and NNC utilised in this work, as well as training strategies and hyperparameters for the NNO and the NN output model are provided in Appendices A and B.

\section{Study cases}
\label{sec:studycases}

This section presents a description of the three systems analysed using the proposed methodology. We first test a modified Kuramoto-Sivashinsky equation to check the observer performance with a simple partial differential equation. The task consists of estimating the complete velocity field using measurements from a reduced number of sensors. In this case, we introduce measurement noise and investigate training both with and without noise addition. 
The second problem studied consists of a confined cylinder flow with small jets used as actuators. By reading signals from sensors located on the walls, the goal is to estimate the velocity values at a set of points along the flow field. For these two first cases, we leverage the estimated states to feedback stabilizing NNCs, which should become possible, even though direct measurements are not available for the states vector.
Finally, we test the approach with PIV data from a turbulent boundary layer experiment. In this case, low-resolution noise-corrupted sensor data is employed to estimate the velocity fields along the boundary layer.

\subsection{Modified Kuramoto-Sivashinsky equation} 
\label{sec:met_ks}

To test the proposed observation methodology, the partial differential equation known as the Kuramoto-Sivashinsky (KS) equation is chosen. It can be written as
\begin{equation}
    \frac{\partial \phi}{\partial t} + \phi\frac{\partial \phi}{\partial x_c} = -\frac{1}{R}\left( P\frac{\partial ^2 \phi}{\partial x_c^2} +\frac{\partial ^4\phi}{\partial x_c^4}\right) \mbox{ ,}
\end{equation}
where $R$ is equivalent to the Reynolds number in a fluid flow, and $P$ represents a balance between energy production and dissipation. The choice $R=0.25$ and $P=0.05$ with periodic boundary conditions corresponds to a chaotic and globally unstable case. The spatial coordinate is represented by $x_c$. The dynamical system solved in this work is a modified version of the KS equation that adds control inputs to actively access the states, besides including a term to ensure that a single uniform natural equilibrium configuration $\phi(x_c) = V$ exists. It can be written as
\begin{equation}
    \frac{\partial \phi}{\partial t} + \phi\frac{\partial \phi}{\partial x_c} = -\frac{1}{R}\left( P\frac{\partial ^2 \phi}{\partial x_c^2} +\frac{\partial ^4\phi}{\partial x_c^4}\right) -\frac{Q}{L} \int_{0}^{L}(\phi(x_c)-V)\,dx_c + \sum_{i=0}^m B_i(x_c) u_i\mbox{ .} \label{eq:ks_mod}
\end{equation}
The term $-Q/L \int_{0}^{L}(\phi(x_c)-V)\,dx_c$ ensures that, for $u(t) = 0$, $\phi(x_c) = V$ is the only possible uniform natural equilibrium possible. We choose $Q=0.0005$, $V=0.2$ and $L=60$, where L is the domain length. With these values, the partial differential equation is globally unstable and presents a chaotic behaviour. The vector $\bm{u}$ is composed of $m=3$ control inputs $u_i$ that modulate the amplitude of 3 evenly spaced Gaussian windows $B_i$ along the spatial domain.

\begin{figure}
\centering
\begin{overpic}[width=0.98\linewidth]{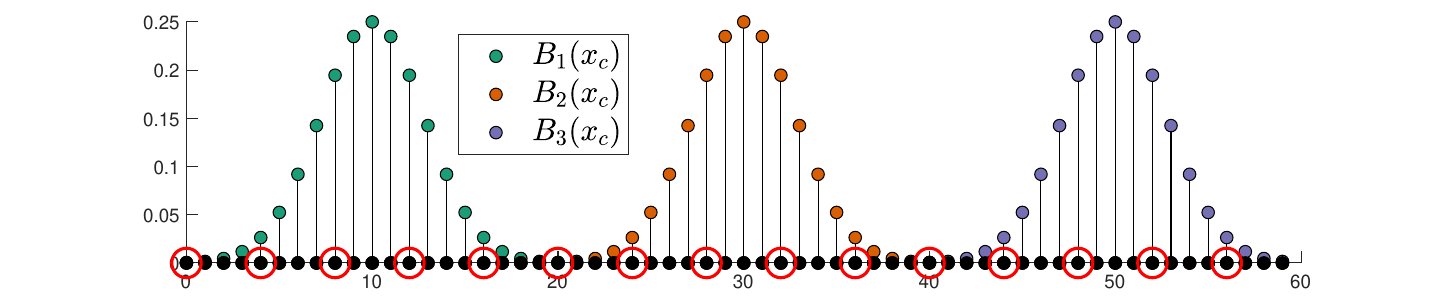}
\end{overpic}
\begin{overpic}[width=0.98\linewidth]{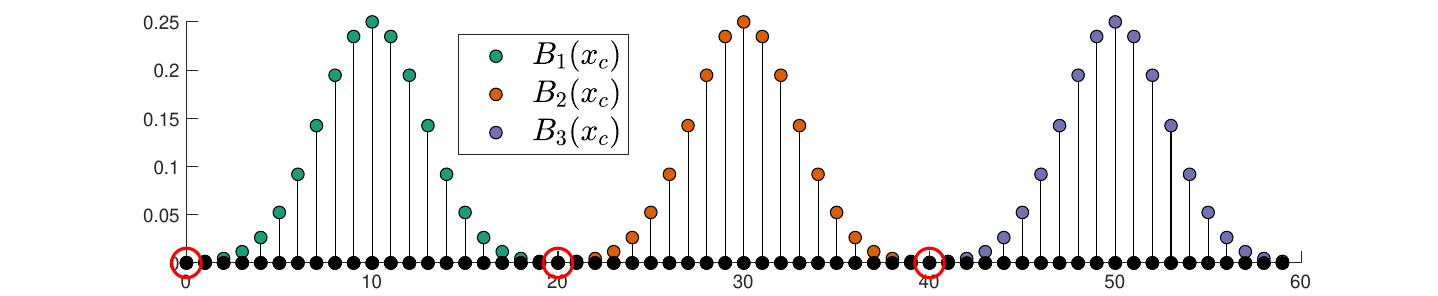}
\end{overpic}
\caption{State, sensor, and actuation scheme for discretized KS equation. The black dots represent each of the 60 state positions. The sensor locations are highlighted with red circles for the cases with 15 (top) and 3 (bottom) sensors. The coloured dots show the Gaussian actuation profiles.}
\label{fig:ks_setup}
\end{figure}

The system is spatially discretized by explicit 4th-order centred finite difference schemes with $\Delta x_c = 1$ and periodic boundary conditions. Evolution in time is performed along a time window described in appendix B for data gathering, using the standard 4th-order Runge-Kutta scheme with $\Delta t = 0.025$, ensuring numerical stability. Control and estimation are conducted at $\Delta t_c=400\Delta t = 10$, the same sampling time of the trained NNSMs, such that $t=k\Delta t_c$, where $t$ and $k$ are the continuous and discrete time variables, respectively --- following control theory notation, the discrete time is represented by natural numbers. From the signals $\phi(x_c)$ at the 60 points that compose $\bm{x} = [\, \phi(0), \dots, \phi(59) \,]^T$, we choose $n_s$ evenly distributed ones to compose the sensors vector $\bm{y}$. Figure \ref{fig:ks_setup} presents the discretized states, outputs and actuation schemes, showing examples where different numbers of sensors are used, whose positions are represented by the red circles. 
In this work, we present results of several study cases with this implementation of the modified KS equation, including tests with either 15 or 3 sensors. We also add white or time-correlated Gaussian measurement noise in some cases, as well as sensing with $\Delta k > 1$.
The nonlinear operators $\widetilde{\mathcal F}(\cdot)$, $\widetilde{\mathcal C}(\cdot)$, ${\mathcal K}(\cdot)$ and ${\mathcal L}(\cdot)$ are implemented as NNs with fully connected hidden layers. The NNSM $\widetilde{\mathcal F}$ and the NNC ${\mathcal K}$ are the same as those trained by \cite{deda2024neural}. At the beginning of the simulation for data sampling, the initial condition $\phi(x_c)=V$ is chosen.

\subsection{Confined cylinder flow} \label{sec:met_cyl}

\begin{figure}
    \centering
    \includegraphics[width=0.98\linewidth]{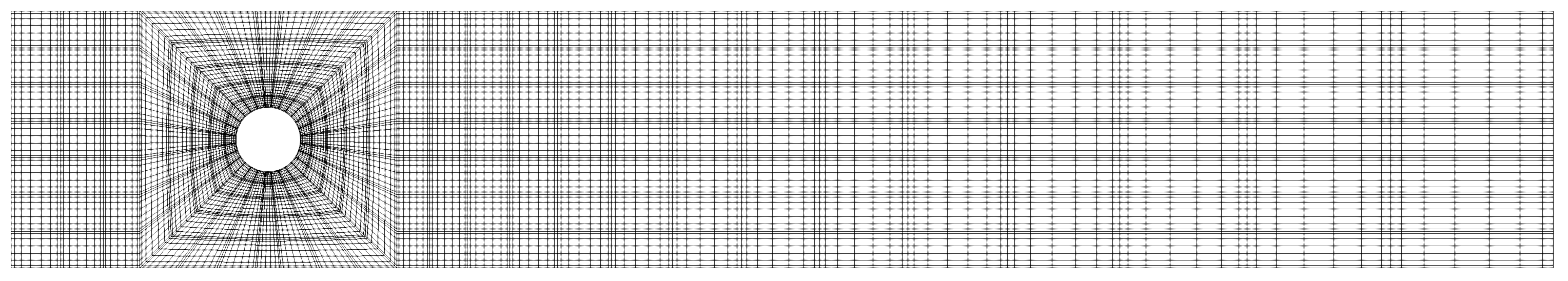}
    \caption{Confined cylinder flow domain and computational grid.}
    \label{fig:cyl_grid}
\end{figure}

In the present work, a confined cylinder flow is also investigated. The setup is implemented in Nek5000 \citep{li2022reinforcement} for a Reynolds number $\mathrm{Re} = 150$, based on the cylinder diameter $D$ and the average inlet velocity. At such conditions, the flow is globally unstable, presenting periodic vortex shedding. The upper and lower channel walls are spaced $H=4D$ apart, with the cylinder centred in between, $4D$ away from the inflow and $20D$ away from the outflow. The employed grid and geometry are presented in figure \ref{fig:cyl_grid}. The simulations are performed using a timestep $\Delta t = 5.0\times10^{-3}$ along the time windows described in appendix B for gathering training data, while sampling for control and estimation is performed at $\Delta t_c=40\Delta t = 2\times10^{-1}$. Again, $t=k\Delta t_c$, where $t$ and $k$ are the continuous and discrete time variables, repectively.

\begin{figure}
\centering
\begin{overpic}[width=.43\linewidth,trim={0cm 0cm 10cm 0cm},clip]{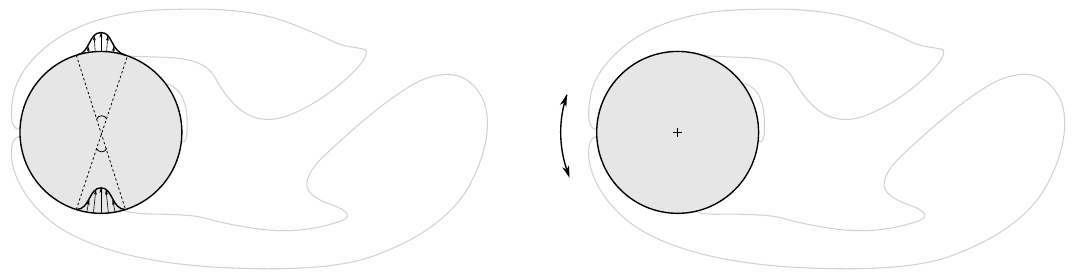}
\put(22,28.5){\tiny{$10\degree$}}
\end{overpic}
\caption{Actuation scheme applied to the cylinder flow. Blowing/suction jets in opposition are modulated by a single control input.}
\label{fig:cyl_act}
\end{figure}

To actively modify the flow, mini-jets are implemented in phase opposition as Dirichlet boundary conditions, providing zero-net mass-flux. The control input $\bm{u}=Q^*=Q/Q_\mathrm{ref}$ consists of a single entry: the normalized injected mass flow rate at a single mini-jet, where
\begin{equation}
    Q_\mathrm{ref} = \int_{-D/2}^{D/2}\phi_\rho\phi_udy \mbox{ .}
\end{equation}
Here, $\phi_\rho=1$ is the density of the incompressible flow and $\phi_u = 6(H/2-y)(H/2+y)/H^2$ is the horizontal component of velocity at the inflow, which is a parabolic profile. A cosine window function along the angular coordinate (coincident with the cylinder centre) sets the profile presented in figure \ref{fig:cyl_act}, such that the maximum absolute values take place at the bottom-most and top-most points of the cylinder wall. The actuation effort is limited to $|Q^*|<0.06$.

\begin{figure}
\centering
\begin{overpic}[width=0.49\linewidth,trim={1.5cm 2cm 2cm 1.8cm},clip]{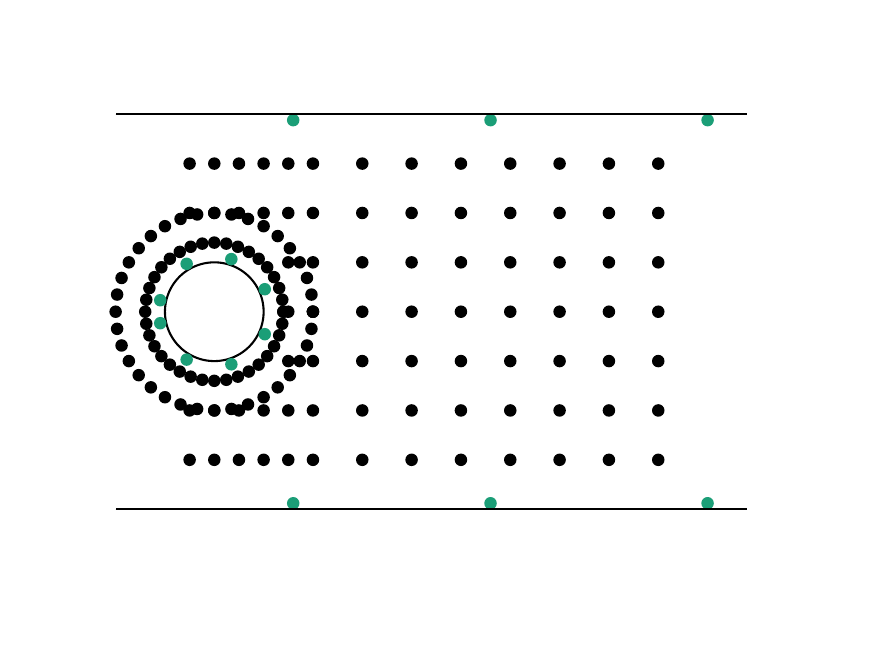}
\end{overpic}
\begin{overpic}[width=0.49\linewidth,trim={1.5cm 2cm 2cm 1.8cm},clip]{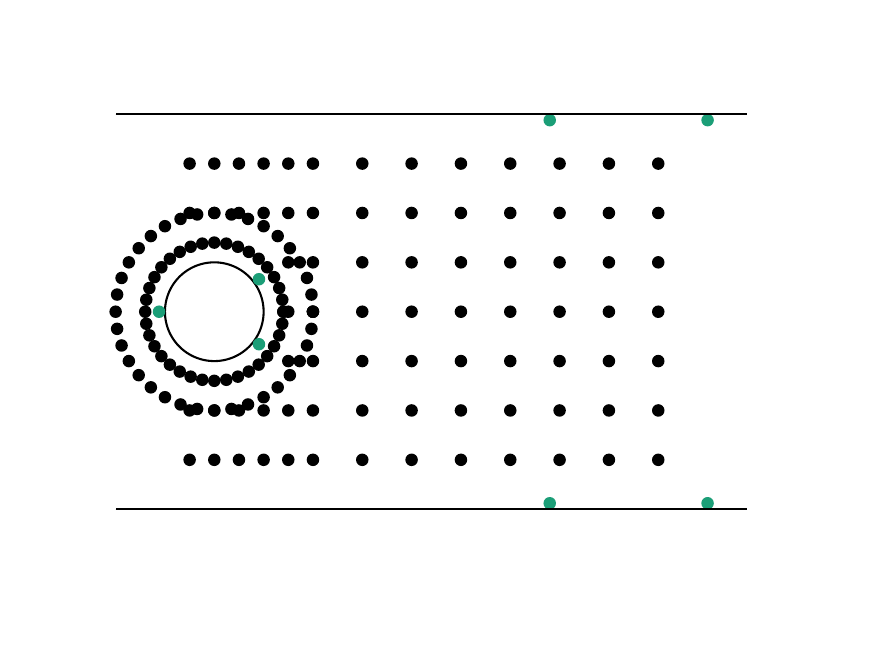}
\end{overpic}
\caption{
Probing setups for the confined cylinder flow, featuring 14 (left) and 7 (right) sensor locations. The state vector $\bm{x}$ consists of horizontal and vertical velocity components at locations indicated by black dots. The measurement vector $\bm{y}$ consists of pressure measurements taken at green dots near the cylinder and channel walls.}
\label{fig:cyl_sensors}
\end{figure}

The vector $\bm{x}$ (featuring 306 states) is composed of the horizontal and vertical velocities $\phi_u$ and $\phi_v$, respectively, at 153 locations marked with black dots in figure \ref{fig:cyl_sensors}. This is the same setup previously studied by \cite{rabault2019artificial}, and the state feedback NNC proposed by \cite{deda2024neural} successfully stabilized this flow configuration. In the present work, we assume that measurements of $\bm{x}$ are unavailable, proposing more realistic sensor setups to compute estimates $\bm{\hat{x}}$, which in turn are fed to the NNC. Here, pressure measurements $\phi_p$ are probed on the wall, at locations marked with green dots in figure \ref{fig:cyl_sensors}. Two setups are analysed, where either 14 or 7 sensors are used to compose $\bm{y}$.
Here, the nonlinear operators $\widetilde{\mathcal F}(\cdot)$, $\widetilde{\mathcal C}(\cdot)$, ${\mathcal K}(\cdot)$ and ${\mathcal L}(\cdot)$ are also implemented as NNs with fully connected hidden layers. The NNSM $\widetilde{\mathcal F}$ and the NNC ${\mathcal K}$ are also provided by \cite{deda2024neural}.

\subsection{Turbulent boundary layer} \label{sec:met_bl}
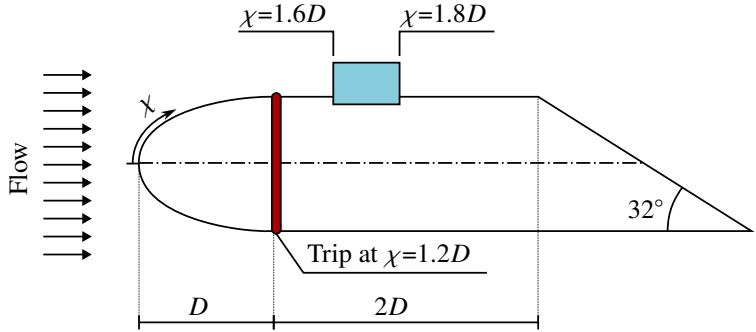
\begin{figure}
    \centering
    \begin{overpic}[width=0.6\linewidth,trim={-0.7cm 0cm 0cm -1.5cm},clip]{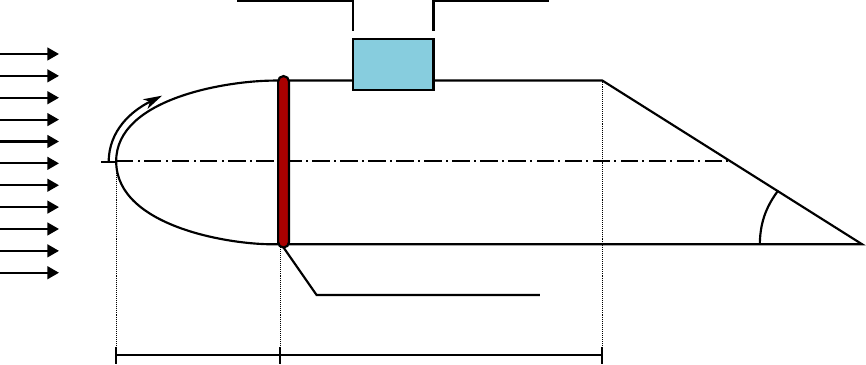}
    \put(24,2){$D$}
    \put(49,2){$2D$}
    \put(0,18){\rotatebox{90}{Flow}}
    \put(17,29){\rotatebox{35}{$\chi$}}
    \put(40,9){Trip at $\chi$=1.2$D$}
    \put(83,15){$32^\circ$}
    \put(31,41.3){$\chi$=1.6$D$}
    \put(52.5,41.3){$\chi$=1.8$D$}
    \end{overpic}
    \caption{Experimental setup of the slanted cylinder flow. Time-resolved PIV images of a turbulent boundary layer are captured within the cyan window.}
    \label{fig:bl_setup}
\end{figure}

The last problem analysed consists of a turbulent boundary layer developing over a bullet-shaped body with a slanted cut. The experimental setup used in this study was originally proposed by \cite{zigunov2020detailed}, who collected PIV data which are used for the present NN estimation trials. The setup is shown in figure \ref{fig:bl_setup}, where $\phi_u$ and $\phi_v$ velocity components are captured within the cyan window. The experiment is performed at Reynolds number $\mathrm{Re} = 40\,000$ relative to diameter $D= \mathrm{146.05 mm}$ of the cylinder section. The freestream velocity is 4.1 m/s, and the measurements are sampled at 5000 frames per second. The dataset provided for the current NN trainings were post processed, providing 1999 snapshots, including the velocity components $\phi_u$ and $\phi_v$, as well as pressure $\phi_p$, obtained via the one-shot omnidirectional pressure integration (OS-MODI) through matrix inversion \citep{zigunov2024one}.

\begin{figure}
    \centering
    \begin{overpic}[width=0.32\linewidth]{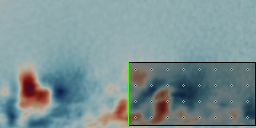}
    \put(2,40){$(a)$}
    \end{overpic}
    \includegraphics[width=0.32\linewidth]{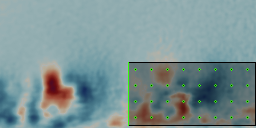}
    \includegraphics[width=0.32\linewidth]{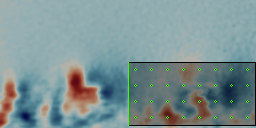}
    \begin{overpic}[width=0.32\linewidth]{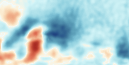}
    \put(2,40){$(b)$}
    \end{overpic}
    \includegraphics[width=0.32\linewidth]{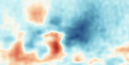}
    \includegraphics[width=0.32\linewidth]{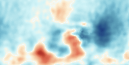}
    \begin{overpic}[width=0.32\linewidth]{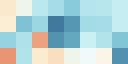}
    \put(2,40){$(c)$}
    \end{overpic}
    \includegraphics[width=0.32\linewidth]{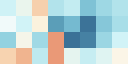}
    \includegraphics[width=0.32\linewidth]{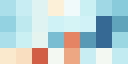}
    \caption{Snapshots depicting contours of $\phi_v$ along the boundary layer. Each frame is 10 time steps apart from their neighbour. The PIV-sampled image is shown in row (a), where the window used to compose $\bm{x}$ and the probe locations (green dots) chosen for the problem are highlighted. The vertical green line shows the region where the boundary conditions $\bm{u}$ are imposed in the model. The $\phi_v$ components of $\bm{x}$ (states) and $\bm{y}$ (probed values) are depicted in rows (b) and (c), respectively.}
    \label{fig:bl_shots}
\end{figure}

Figure \ref{fig:bl_shots} (a) presents contours of $\phi_v$ at snapshots spaced 10 time steps apart so the convection of structures can be clearly noticed. To compose the states vector $\bm{x}$, we use the smaller window indicated in the the figure, resulting in two images containing $128\times64$ pixels each, one for $\phi_u$ and one for $\phi_v$ --- the latter being depicted in figure \ref{fig:bl_shots} (b) --- totalling $128\cdot64\cdot2 = 16\,384$ states. At the measured location, the boundary layer is turbulent due to bypass transition from tripping implemented in the experiment; the source of disturbances is not contained within the data. This brings a relevant issue when modelling the flow dynamics, as it is not possible to predict the next time step without having information from upstream. To solve this issue, we use the control inputs vector $\bm{u}$ to represent the boundary conditions, assuming them as the source term that affects the states. This source term is composed by 32 values of $\phi_u$ plus 32 of $\phi_v$ probed at the 
green line shown in figure \ref{fig:bl_shots} (a), where measurements at every second pixel are taken, totalling $64$ entries for $\bm{u}$. Although this is not a realistic approach for real world applications since it would require real-time measurements in many locations away from the walls, it is a workaround to test the NNO capabilities when dealing with complex turbulent flows. In real applications, the region encompassing the states could include the position where the disturbances arise, and therefore such forcing modelling would not be required. Alternatively, hot-wire sensors could be implemented to measure flow velocities away from the wall, combined with other types of NN architectures, such as Graph NNs, which do not require regular grids like the CNNs. 
Finally, to compose the outputs vector $\bm{y}$, we employ $\phi_u$ and $\phi_v$ values at the $4\times8$ array green dots shown in figure \ref{fig:bl_shots} (a), totalling 64 variables. Figure \ref{fig:bl_shots} (c) presents the resulting low resolution image built from the readings at these sensor locations.

\begin{figure}
    \centering
    \begin{overpic}[width=0.98\linewidth,trim={.3cm 0cm 1cm 0cm},clip]{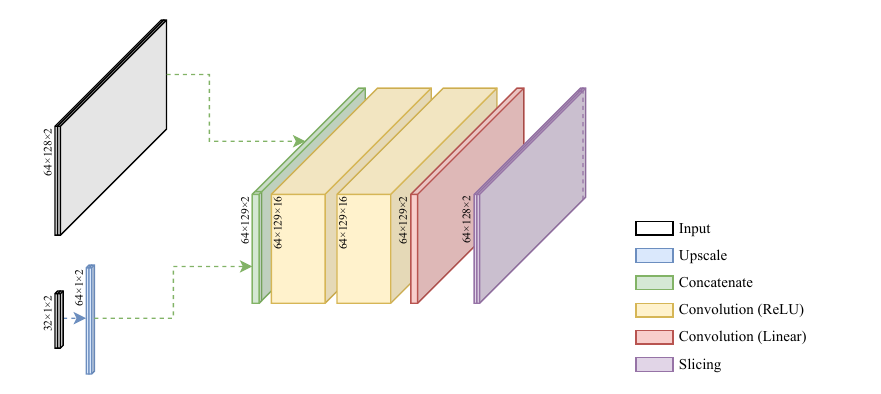}
    \put(0,7.){\scalebox{.8}{\rotatebox{90}{$\bm{u}[k]$}}} 
    \put(0,24){\scalebox{.8}{\rotatebox{90}{$\bm{x}[k]$}}} 
    \put(72,28){\scalebox{.8}{\rotatebox{90}{$\bm{x}[k+1]$}}} 
    \end{overpic}
    \caption{Neural network architecture used for the turbulent boundary layer NNSM. ReLU activation is used, except for the output convolution layers, which are linear.}
    \label{fig:bl_nnsm}
\end{figure}

A specialized NN architecture is developed to process the high number of states. As presented in figure \ref{fig:bl_nnsm}, $\phi_u$ and $\phi_v$ are concatenated to form a 2-channel image. The same is done to $\bm{u}$, which is upscaled through a nearest-neighbour algorithm. The states and boundary conditions are combined to take advantage of the spatial relations between them. Pure convolutions are applied in succession, such that pixels are updated only based on their neighbours. By avoiding fully connected layers, we also take better advantage of the limited dataset size, since each pixel of $\bm{x}$ can be seen as a data unit that goes through the same nonlinear function. At the output, the resulting image is sliced so the initial size is recovered. 
%
%
Each component $\phi_u$ and $\phi_v$ is normalized separately so that their values range from zero to one.
This normalization process is performed for all the networks trained, and the input ($\bm{u}$) and output ($\bm{y}$) vectors are slices of these normalized variables.
 
\begin{figure}
    \centering
    \begin{overpic}[width=0.98\linewidth,trim={0.2cm 0cm .9cm 0cm},clip]{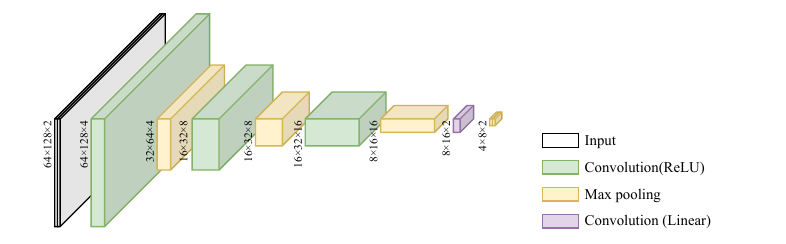}
    \put(1,7){\scalebox{.8}{\rotatebox{90}{$\bm{x}[k]$}}} 
    \put(69.1,16.2){\scalebox{0.8}{\rotatebox{90}{$\bm{y}[k]$}}} 
    \end{overpic}
    \caption{Neural network architecture used for the turbulent boundary layer NN output model. ReLU activation is used, except for the output layer, which is linear.}
    \label{fig:bl_nnout}
\end{figure}

\begin{figure}
    \centering
    \begin{overpic}[width=0.98\linewidth,trim={0.6cm 0cm .9cm 0cm},clip]{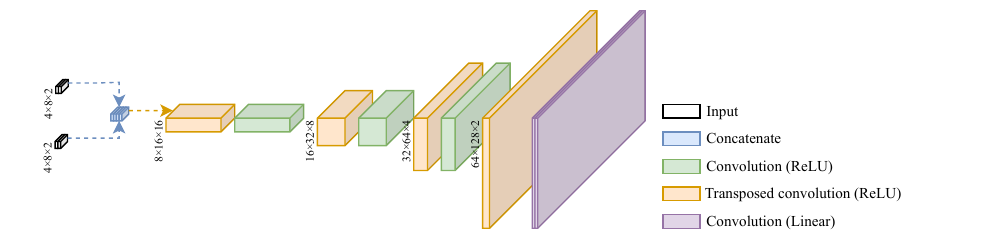}
    \put(0,3.){\scalebox{.8}{\rotatebox{90}{$\hat{\bm{y}}[k]$}}} 
    \put(0,18){\scalebox{.8}{\rotatebox{90}{$\bm{y}[k]$}}} 
    \put(68.5,17){\scalebox{.8}{\rotatebox{90}{$\bm{v}[k]$}}} 
    \end{overpic}
    \caption{Neural network architecture used for the turbulent boundary layer NNO. ReLU activation is used, except for the last fully-connected layer, which is linear.}
    \label{fig:bl_nno}
\end{figure}

The architecture for the output model is shown in figure \ref{fig:bl_nnout}. Here, the input states are downscaled through convolution and max pooling steps. The outputs correspond to the low resolution images shown in Figure \ref{fig:bl_shots} (c). Although this network could be implemented as a simple slicing function --- thus not requiring trainable parameters --- we implement it as a nonlinear function to avoid the assumption that $\mathcal{C}(\bm{x})$ is known. Finally, the NNO architecture is presented in figure \ref{fig:bl_nno}, where the correction signal $\bm{v}[k]$ consists of the 2-channel output image, which is summed to the corresponding predicted state images.

Several limitations of the proposed methodological approach should be noted. The sensors used are limited to the plane where variables were sampled during the experiments, which can hinder turbulence modelling given its three-dimensional nature, particularly considering the absence of a third velocity component ($\phi_w$) in the dataset. Furthermore, since the proposed experiment was not configured for active real-time control and considering that our group lacks experimental resources to reproduce the proposed experiment, the deployment of the NNOs is limited to open-loop tests, where control input signals adopt the role of the aforementioned boundary conditions contained within data, propagated along time through the NNSM, whose predictions are rectified by the observer. 

To illustrate the advantage of implementing the proposed estimation topology over the reconstruction of $\bm{x}$ from $\bm{y}$ through direct NN inference, we also train an NN super-resolution model to upscale the low-resolution images into the original states. The architecture employed is similar to that presented for the NNO (see figure \ref{fig:bl_nno}) for a fair comparison. The difference is that the $\hat{\bm{y}}$ input to the NNO and the concatenation layer are skipped, since there is no dynamic model of the flow to leverage predictions. Additionally, the super-resolution NN outputs the reconstructed states $\bm{x}$ directly instead of $\bm{v}$. The NNO and the super-resolution NN are trained to estimate the states in the presence of white Gaussian noise with $\sigma = 0.06$, which is artificially added to the normalized values of $\phi_u$ and $\phi_v$. Although the experimental data is already contaminated with unknown measurement noise, we further add the artificial source to better test the NNO ability to work under non-ideal conditions.

Since the study is conducted for a small spatial window, the problem is dominated by convection, although some distortions are seen in the flow structures being transported. Due to these characteristics, we also implement a convective model, which is built by assuming a frozen-field (FF) hypothesis. We take the boundary conditions vector $\bm{u}$ and apply the transport equations
\begin{align}
    \frac{\partial\phi_u}{\partial t} &= -\bar{\phi}_u(x,y)\frac{\partial\phi_u}{\partial x} \mbox{ ,} \\
    \frac{\partial\phi_v}{\partial t} &= -\bar{\phi}_u(x,y)\frac{\partial\phi_v}{\partial x} \mbox{ ,}
\end{align}
where the mean horizontal flow velocity $\bar{\phi}_u(x,y)$ is computed at each pixel. The spatial discretization is performed through a 1st-order backward finite difference scheme, and time integration is conducted using the 4th order Runge-Kutta scheme. 
To ensure numerical stability, the timestep is reduced to one fifth of the original $\Delta t$ at which data was sampled, although the results are only shown at the time instants that match the experiment. Since the boundary conditions have half the resolution of the final image columns, we simply replicate each pixel when upscaling.

\section{Results}
\label{sec:results}

\begin{figure}
\centering
\begin{overpic}[width=0.9\linewidth]{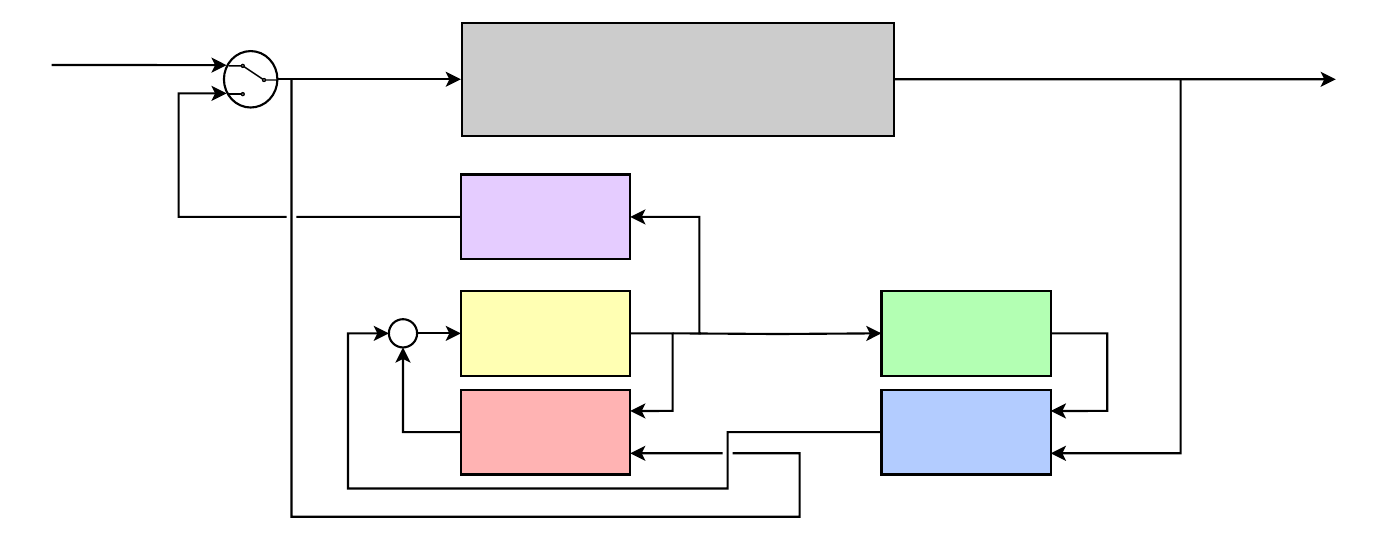}
\put(7,35.6){\scalebox{1.0}{$\bm{u}_\mathrm{o}[k]$}} 
\put(25,24.7){\scalebox{1.0}{$\bm{u}_\mathrm{c}[k]$}} 
\put(25,34.6){\scalebox{1.0}{$\bm{u}[k]$}} 
\put(54,16.3){\scalebox{1.0}{$\hat{\bm{x}}[k]$}} 
\put(77,34.6){\scalebox{1.0}{$\bm{y}[k]$}} 
\put(77,16.3){\scalebox{1.0}{$\hat{\bm{y}}[k]$}} 
\put(54,9.2){\scalebox{1.0}{$\bm{v}[k]$}} 
\put(40.7,33){\scalebox{1.0}{Nonlinear plant}} 
\put(37.2,14.5){\scalebox{1.0}{$z^{-1}$}} 
\put(36.7,22.9){\scalebox{1.0}{$\mathcal{K}(\cdot)$}} 
\put(36.7,7.4){\scalebox{1.0}{$\widetilde{\mathcal{F}}(\cdot)$}} 
\put(67,7.4){\scalebox{1.0}{$\mathcal{L}(\cdot)$}}
\put(67,14.5){\scalebox{1.0}{$\widetilde{\mathcal{C}}(\cdot)$}}     
\end{overpic}
\caption{Block diagram representing the nonlinear observer in closed-loop to estimate flow states. The control input can be toggled between open-loop and closed-loop. The nonlinear plant returns $\bm{y}[k]$ as a function of $\bm{u}[k]$ and its internal states.}
\label{fig:loop_deployment}
\end{figure}
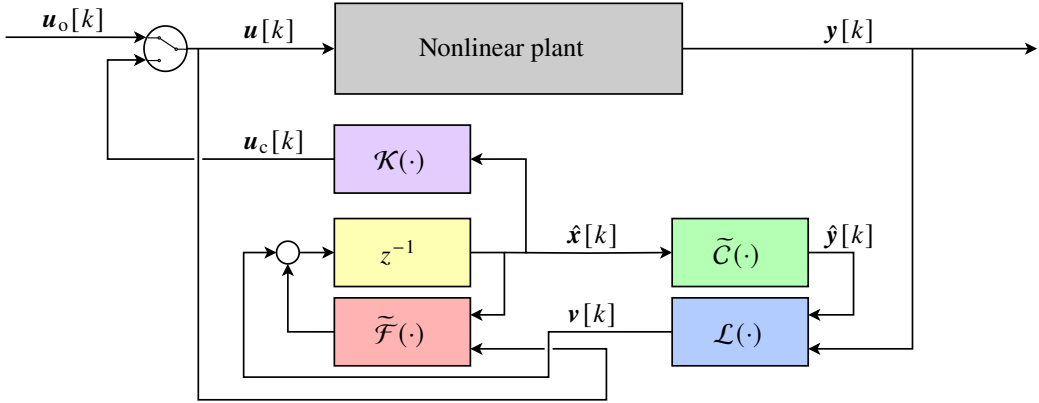

In this section, results are presented for the study cases, where the trained NNOs are deployed in closed loop as illustrated in figure \ref{fig:loop_deployment}. For the KS equation and the cylinder flow, the nonlinear plant is a numerical simulator, which is run to produce the results presented in the current section. For these cases, we present both open-loop (for illustrative purposes) and closed-loop control results. For open-loop tests, the system is perturbed using staircase signals as control inputs ($\bm{u}=\bm{u}_\mathrm{o}$). For closed-loop control, the stabilizing NNC is incorporated into the loop, fed solely by the estimated states, i.e., $\bm{u}=\bm{u}_\mathrm{c}=\mathcal{K}(\hat{\bm{x}})$. The states vector $\bm x$ are internal to the nonlinear plant, and therefore unknown to the control/observer loops. The initial condition is the average state in the training dataset
\begin{equation}
    \hat{\bm x}[0] = \frac{1}{n_t}\sum_{i=1}^{n_t} \bm{x}_i \mbox{ ,}
\end{equation}
where $n_t$ is the total number of samples.

For the experimental boundary layer setup, only legacy data from prior experiments are available. 
We only conduct open-loop trials by using the control inputs ($\phi_u$ and $\phi_v$ boundary conditions) provided in the dataset. Therefore, the nonlinear plant is reduced to accessing the dataset at each time step. The loop structure, however, is still the same shown in figure \ref{fig:loop_deployment}, but $\bm{u}[k]$ always receives the prescribed input signal $\bm{u}_\mathrm{o}[k]$ without any subsequent adjustment for closed-loop control. The initial condition $\hat{\bm x}[0]$ consists of two images --- one for $\phi_u[k=0]$ and one for $\phi_v[k=0]$ --- where each pixel assumes a random value from a uniform distribution.

\subsection{Modified Kuramoto-Sivashinsky equation}

\begin{figure}
\centering
\begin{overpic}[width=0.98\linewidth]{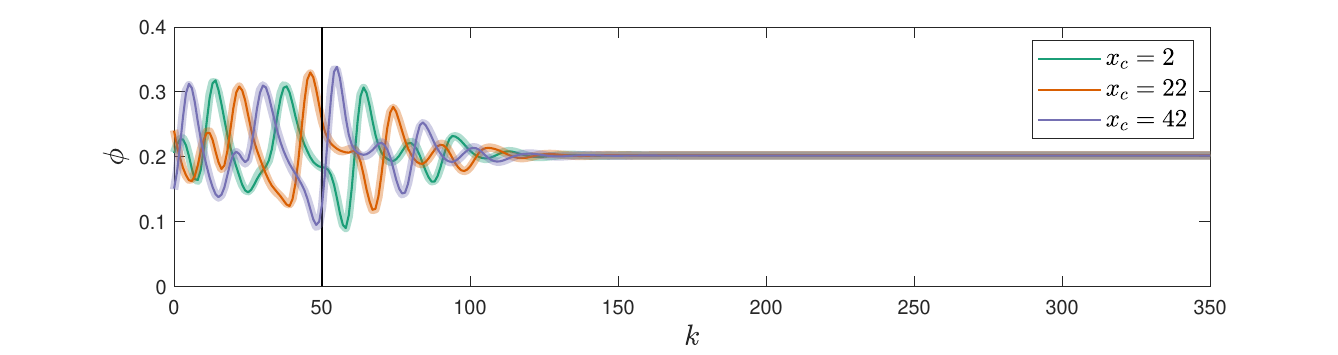}
\end{overpic}
\caption{
Results with 15 sensors for state estimation and feedback control of the modified KS equation. Three states are shown comparing the estimated states $\hat{\bm{x}}$ (thin opaque lines) and the actual states $\bm{x}$ (thick transparent lines), at specific locations $x_c$. Closed-loop control is first applied at $k=50$, as indicated by the vertical black line.}
\label{fig:ks_15sensors}
\end{figure}

\begin{figure}
\centering
\begin{overpic}[width=0.98\linewidth]{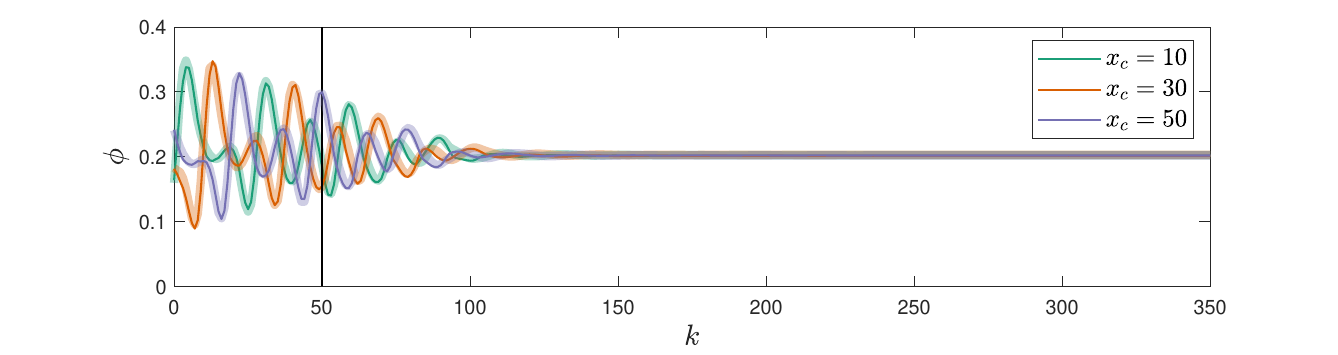}
\end{overpic}
\caption{
As in figure \ref{fig:ks_15sensors}, but with 3 sensors.}
\label{fig:ks_3sensors}
\end{figure}

Results of seven KS simulations are shown following the setup described in \S\ref{sec:met_ks}. A time window of 350 time steps of the simulation are presented, where state tracking under open-loop perturbations is presented for $0\leq k<50$ and with closed-loop control for $50\leq k<350$. We choose to present the last 350 steps, showing a part of the solution that continued from the chaotic attractor; therefore the initial growth from equilibrium is omitted. Figure \ref{fig:ks_15sensors} shows results when 15 sensors are used. The estimated $\phi$ values are presented at coordinates $x_c=2$, $x_c=22$ and $x_c=42$, where no sensors are present. Similarly, figure \ref{fig:ks_3sensors} shows results with only 3 sensors, also at coordinates without sensors. The comparison of the estimated states (thin opaque lines) with the actual states (thicker transparent lines), show that the tracking is almost perfect, with a slight 
difference in the case with fewer sensors. When closed-loop control is turned on, the system is properly stabilized through feedback of the estimated states, with estimation remaining accurate as the system stabilises.  
These first two cases presented, with sets of either 3 or 15 ideal sensors, produce control responses that are very close to those found by \cite{deda2024neural}, who used ideal state feedback. This is expected since the state estimation is very accurate and the same NNC is used.

\begin{figure}
\centering
\begin{overpic}[width=0.98\linewidth]{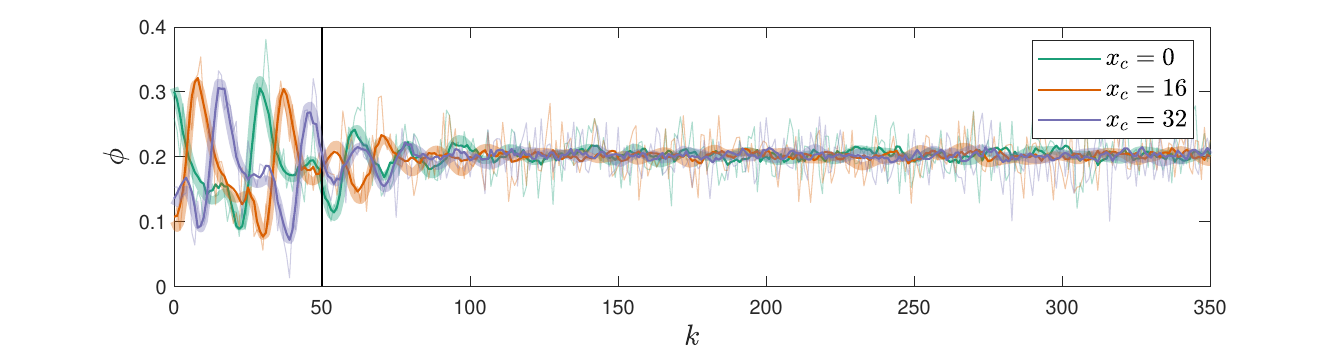}
\end{overpic}
\caption{
Results with 15 sensors and white Gaussian noise with $\sigma = 0.03$ for the modified KS equation. Three states are shown at locations $x_c$ where sensors are present for comparing the estimated states $\hat{\bm{x}}$ (thin opaque lines) and the actual states $\bm{x}$ (thick transparent lines). The thin transparent lines show the measured signals with noise $y+y_n$.}
\label{fig:ks_wnoise}
\end{figure}

\begin{figure}
\centering
\begin{overpic}[width=0.98\linewidth]{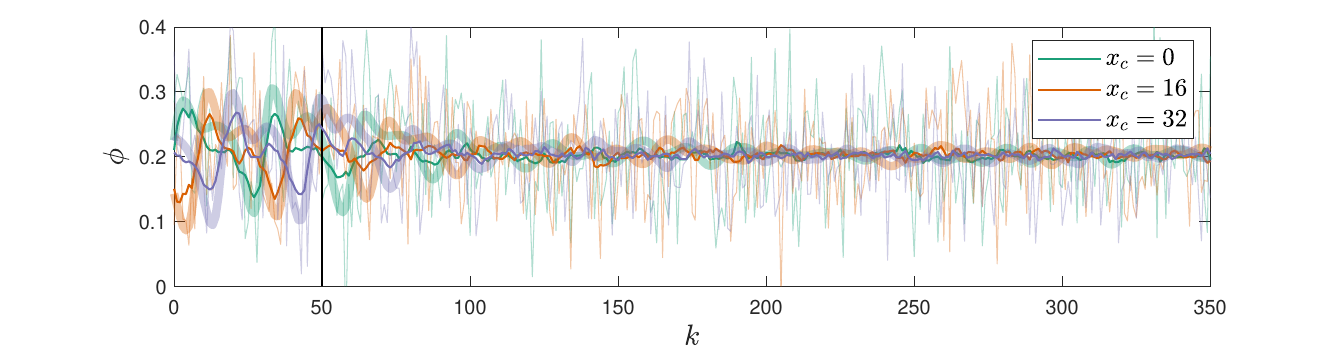}
\end{overpic}
\caption{
As in figure \ref{fig:ks_wnoise}, but with $\sigma=0.07$.}
\label{fig:ks_wnoiselarge}
\end{figure}

Results with added noise are shown in figure \ref{fig:ks_wnoise} for the setup with 15 sensors. Here, the measurement is contaminated with white Gaussian noise with standard deviation $\sigma=0.03$. In this case, states are shown at locations where sensors are present, so a comparison between the actual states and the measured signals (thin transparent lines) can be established. During the open-loop stage, an average signal-to-noise ratio (SNR) of approximately $3.72\times 10^{+0}$ is observed. As the system is controlled, the states oscillations are reduced, lowering the average SNR to around $3.01\times 10^{-2}$. With such low SNR, the observer is not able to estimate the states correctly, and a new range of oscillation amplitude is reached, at which further attenuations are unlikely to occur. In this situation, the controller is not able to bring states to steady-state, but the oscillation amplitudes are reduced to 9.5\% of the original amplitude of the uncontrolled system (measured through the square root of the ratio between SNRs). With $\sigma=0.07$ (see figure \ref{fig:ks_wnoiselarge}), an average open-loop SNR of $7.13\times 10^{-1}$ is found. The controller reduces oscillation amplitudes to 12.5\%, after which an SNR close to $1.10\times 10^{-2}$ is seen. From the plots, it can be noticed that the estimated signals are typically closer to the actual states than the directly measured noise-corrupted signals (thin transparent lines). 

\begin{figure}
\centering
\begin{overpic}[width=0.98\linewidth]{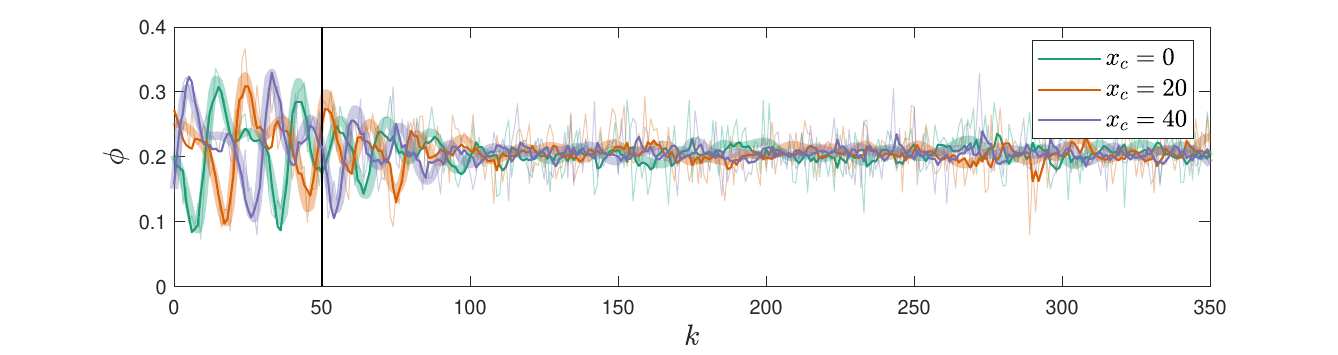}
\end{overpic}
\caption{
As in figure \ref{fig:ks_wnoise}, but with 3 sensors.}
\label{fig:ks_wnoise3sensors}
\end{figure}

With only three sensors and $\sigma = 0.03$ (see figure \ref{fig:ks_wnoise3sensors}), we can also preserve the ability to perform closed-loop control. An amplitude attenuation of $13.9\%$ is seen for the actual states, after which the average SNR is measured at $6.75\times 10^{-2}$. This greater value shows that fewer sensors are worse at observing states under noisy conditions. At the same SNR levels, the other cases studies were still able to estimate states with sufficient accuracy to further attenuate oscillations.

\begin{figure}
\centering
\begin{overpic}[width=0.98\linewidth]{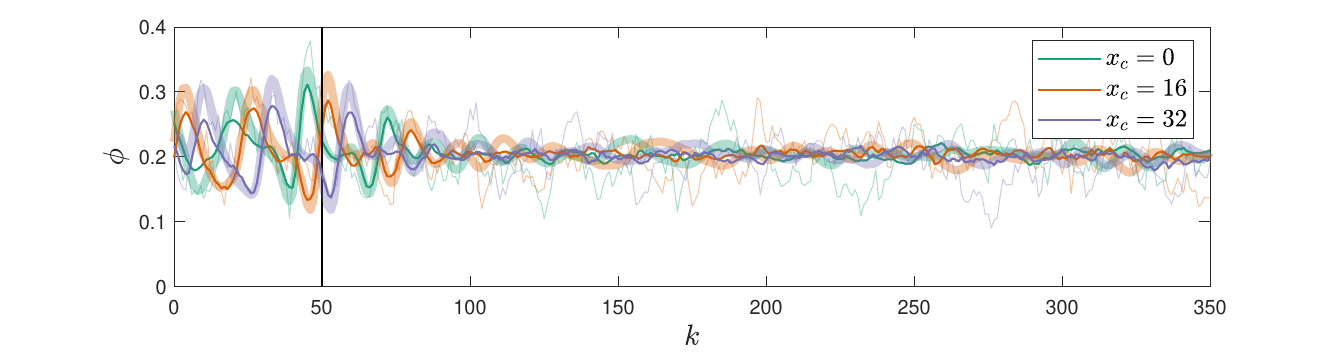}
\end{overpic}
\caption{
As in figure \ref{fig:ks_wnoise}, but with time-correlated noise with $\beta=0.8$.}
\label{fig:ks_cnoise}
\end{figure}

The proposed technique also allows for training with different types of noise. Figure \ref{fig:ks_cnoise} shows a case where time-correlated noise with $\sigma = 0.03$ and $\beta=0.8$ (as defined in \eqref{eq:corrnoise}) is added, both during training and deployment. In this example, using time-correlated noise made attenuation worse, but the NNC is still able to reduce oscillations to around 15.9\% of the original amplitudes, at an average SNR of $9.06\times 10^{-2}$. Since the proposed time-correlated noise presents slower variations along time, it is possible that improved attenuation could be achieved with larger $n_h$ during training. The attenuation values reported here are summarized in table \ref{tab:noise_summary}.

\begin{table}
  \begin{center}
\def~{\hphantom{0}}
  \begin{tabular}{lccc}
    Case & Closed-loop SNR & Final amplitude\\[3pt]
    15 sensors, $\sigma=0.03$, $\beta=0$ & $3.01\times 10^{-2}$ & 9.5\% \\
    15 sensors, $\sigma=0.07$, $\beta=0$ & $1.10\times 10^{-2}$ & 12.5\% \\
    3 sensors, $\sigma=0.03$, $\beta=0$  & $6.75\times 10^{-2}$ & 13.9\% \\
    15 sensors, $\sigma=0.03$, $\beta=0.8$ & $9.06\times 10^{-2}$ & 15.9\% \\
  \end{tabular}
  \caption{Results for each KS case with measurement noise. Final amplitude is measured as the percentage of the uncontrolled oscillation amplitude.}
  \label{tab:noise_summary}
  \end{center}
\end{table}

\begin{figure}
\centering
\begin{overpic}[width=0.98\linewidth]{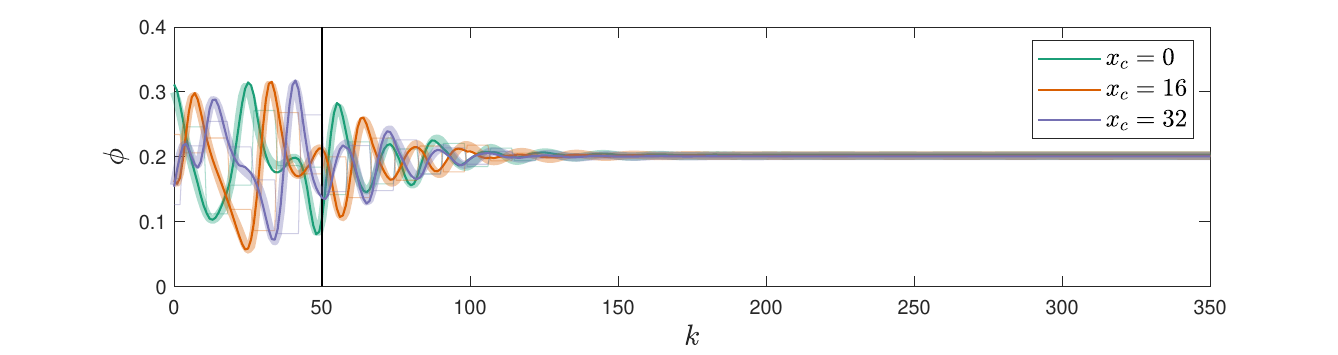}
\end{overpic}
\caption{
Results with 15 sensors and measurements with $\Delta k = 8$ for the modified KS equation. Estimated states $\hat{\bm{x}}$ (thin opaque lines) and the actual states $\bm{x}$ (thick transparent lines) are shown. The staircase signal represents the sensor measurements at a lower sampling rate.}
\label{fig:ks_skip8}
\end{figure}

\begin{figure}
\centering
\begin{overpic}[width=0.98\linewidth]{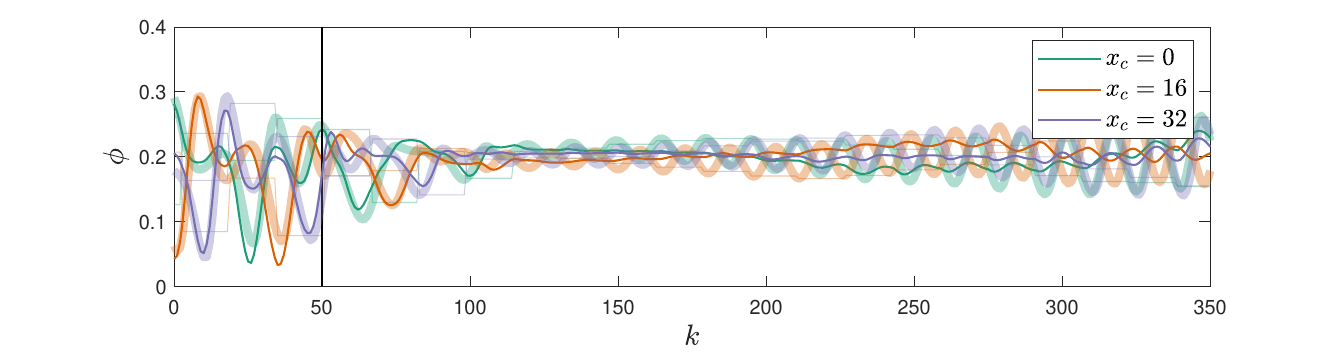}
\end{overpic}
\caption{
As in figure \ref{fig:ks_skip8}, but with $\Delta k = 16$.}
\label{fig:ks_skip16}
\end{figure}

The last two KS cases investigated utilise 15 sensors with either $\Delta k=8$ or $\Delta k=16$, i.e., real-time measurements are only sampled every 8 or 16 time steps of the NNSM/NNC. Thus, for most of the time, the same values $v[k=N\Delta k]$ (where $N \in\mathbb{N}$) are used to correct the NNSM predictions until the next measurement update at $k=(N+1)\Delta k$. With $\Delta k = 8$ (see figure \ref{fig:ks_skip8}), the tracking is nearly perfect, rendering the NNC able to correctly stabilize the plant. In Figure \ref{fig:ks_skip16}, however, we see that $\Delta k = 16$ is enough to prohibit stabilization with the available NNSM. The maximum $\Delta k$ value is probably related to the model ability to produce good predictions for longer, without the need to corrections via sensor information. Therefore, in situations where the output sample period are shorter than the timescales of the error, we should expect the estimation to work well, even with $\Delta k > 1$. Movie 1, submitted as supplementary material, summarizes the time response comparing all cases.

\subsection{Confined cylinder flow}

The validation of closed-loop stabilization of the cylinder flow introduced in \S\ref{sec:met_cyl} is presented here, where the NNO estimates relevant flow variables by reading limited sensor information. Results are shown with actuation performed in two stages: under open-loop perturbations for $0\leq k<200$; and with closed-loop control for $200\leq k<500$. 

\begin{figure}
\centering
\includegraphics[width=0.98\linewidth,trim={3.5cm .6cm 3.5cm .1cm},clip]{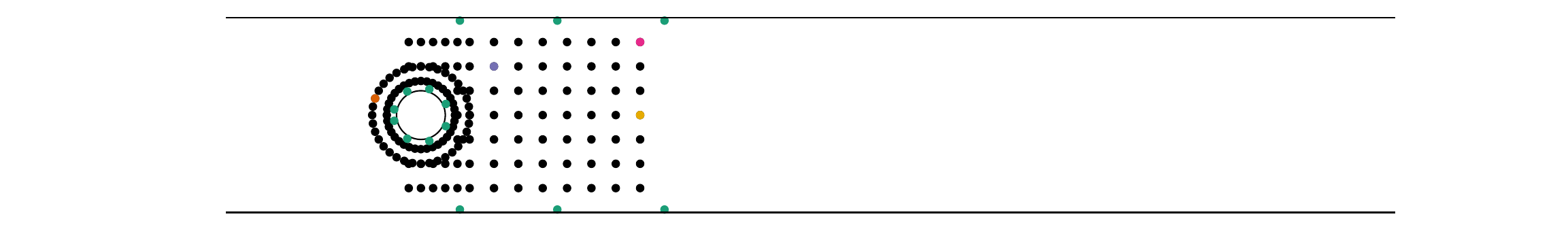}
\includegraphics[width=0.98\linewidth,trim={1.9cm 0.4cm 2cm 0cm},clip]{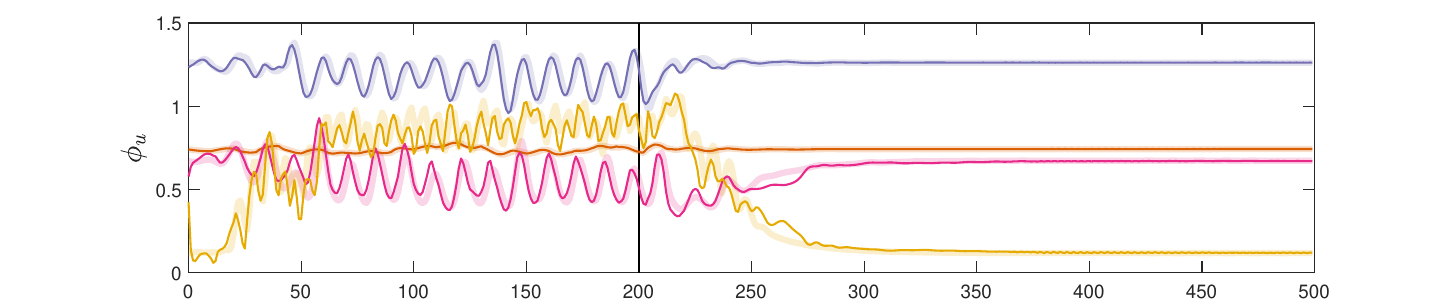}
\includegraphics[width=0.98\linewidth,trim={1.9cm 0cm 2cm 0cm},clip]{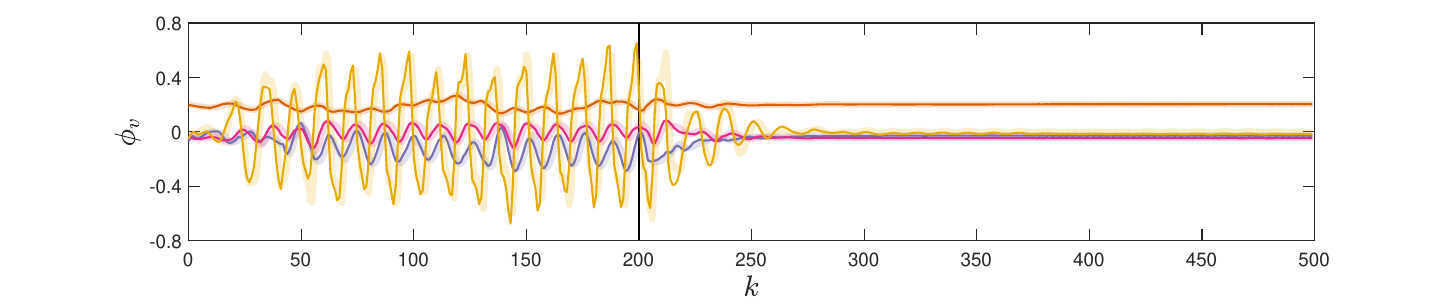}
\caption{
Results for state estimation and control of the confined cylinder flow with 14 sensors.
State curves are shown with the same colours as the respective coloured point locations. Thin opaque and thick transparent lines show estimated and real states, respectively. Closed-loop control is applied at $k=200$, as indicated by the vertical black lines.}
\label{fig:14sensors}
\end{figure}

Results of the case with 14 sensors are presented in figure \ref{fig:14sensors}, which shows $\phi_u$ and $\phi_v$ at a few selected points, marked with different colours along the cylinder wake and upstream of the cylinder. The initial conditions are close to the equilibrium point, and the flow develops to near the limit cycle after a few time steps, before the controller is turned on. Estimated states are shown as thin opaque lines, with the same colours used to represent their respective points in space. By comparing them with the actual states, represented by thicker transparent lines, it is possible to verify that the main tendencies are well captured, specially at the lower shedding frequency that matches lift oscillations. In general, we observe that signals with twice the shedding frequency, related to drag oscillations, are harder to track precisely. This is illustrated by looking at $\phi_u$ in the yellow point, situated at the channel centreline. During the closed-loop stage, the NNO is capable of providing estimations with enough accuracy to enable proper NNC performance. Indeed, only very small oscillations are seen after the main transient, as vortex shedding is almost completely suppressed. The results with 14 sensors are very close to those obtained through direct state feedback (assuming states are known) obtained by \cite{deda2024neural}, albeit slight differences in the settling time occur due to imperfect estimation of states.

\begin{figure}
\centering
\includegraphics[width=0.98\linewidth,trim={3.5cm .6cm 3.5cm .1cm},clip]{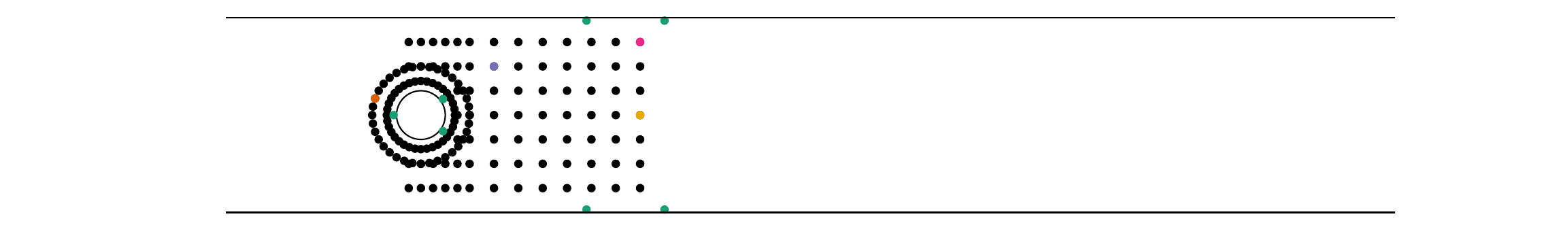}
\includegraphics[width=0.98\linewidth,trim={2.0cm 0.4cm 2cm 0cm},clip]{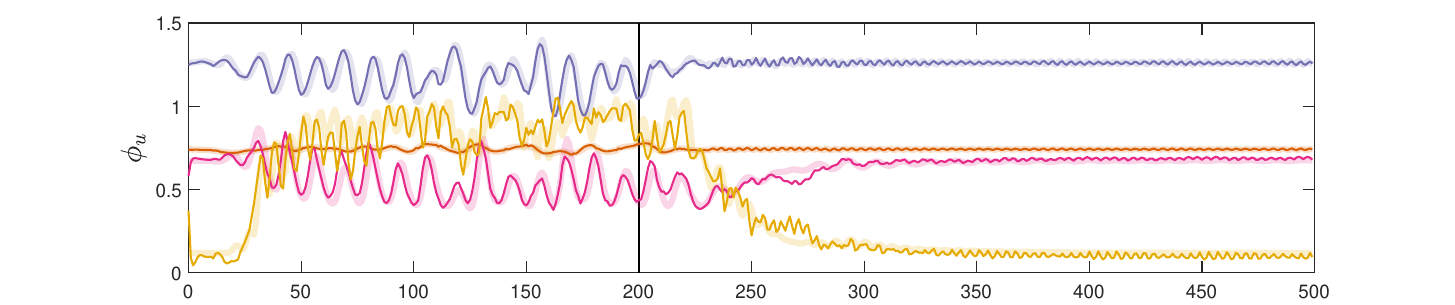}
\includegraphics[width=0.98\linewidth,trim={1.9cm 0cm 2cm 0cm},clip]{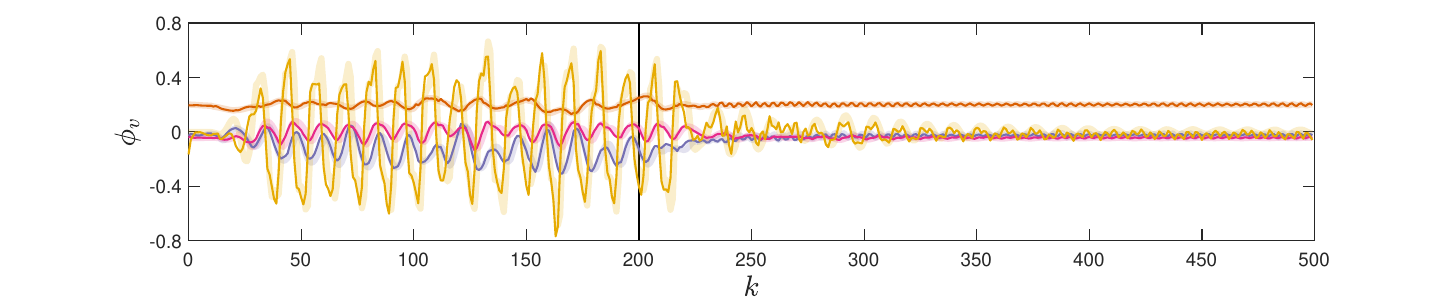}
\caption{
As in figure \ref{fig:14sensors}, but with 7 sensors.}
\label{fig:7sensors}
\end{figure}

\begin{figure}
\centering
\begin{overpic}[width=0.93\linewidth,trim={1.2cm 0.8cm 1.2cm 0.8cm},clip]{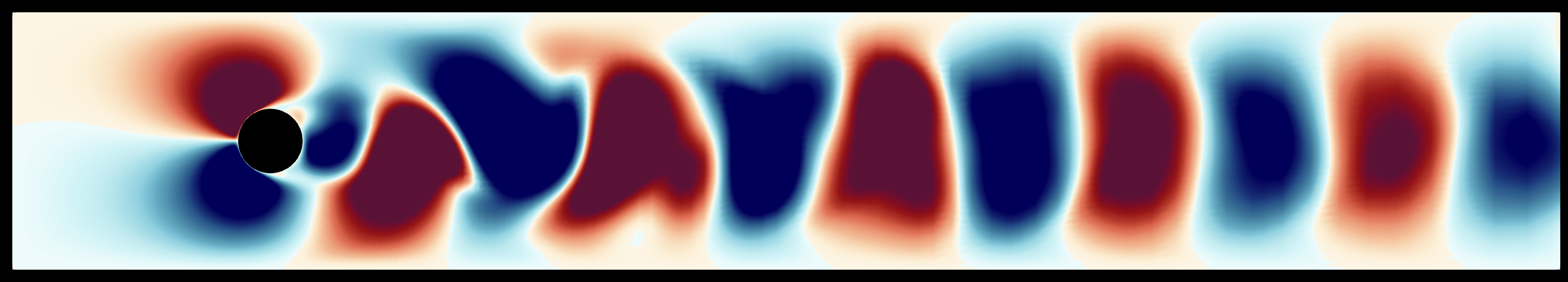}
\put(2,7.5){\textcolor{black}{\scalebox{0.85}{(a)}}}
\end{overpic}%
\begin{overpic}[width=0.049\linewidth,trim={0cm 0cm 0cm 0cm},clip]{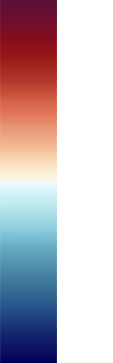}
\put(17,1){\scalebox{0.57}{-0.2}}
\put(17,94){\scalebox{0.57}{+0.2}}
\end{overpic}%
\vspace{.1cm}
\begin{overpic}[width=0.93\linewidth,trim={1.2cm 0.8cm 1.2cm 0.8cm},clip]{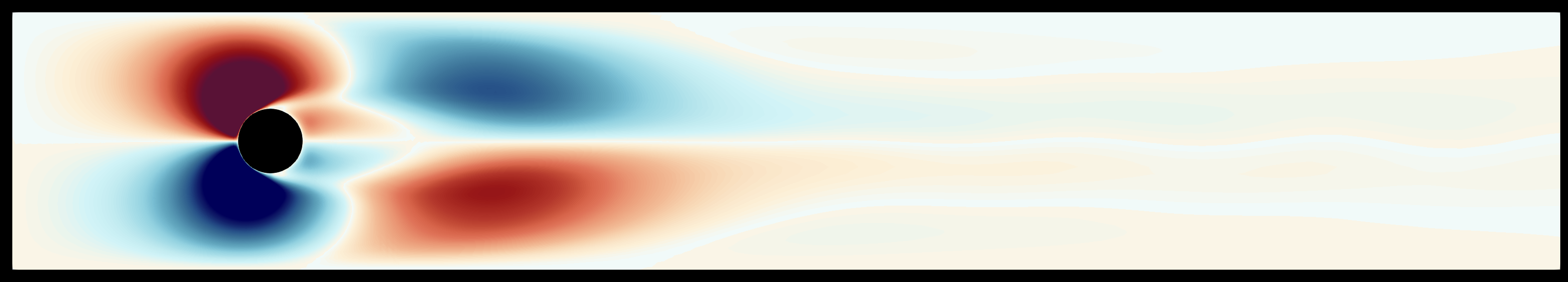}
\put(2,7.5){\textcolor{black}{\scalebox{0.85}{(b)}}}
\end{overpic}%
\begin{overpic}[width=0.049\linewidth,trim={0cm 0cm 0cm 0cm},clip]{figs/cyl_results/scale.png}
\put(17,1){\scalebox{0.57}{-0.2}}
\put(17,94){\scalebox{0.57}{+0.2}}
\end{overpic}%
\vspace{.1cm}
\begin{overpic}[width=0.93\linewidth,trim={1.2cm 0.8cm 1.2cm 0.8cm},clip]{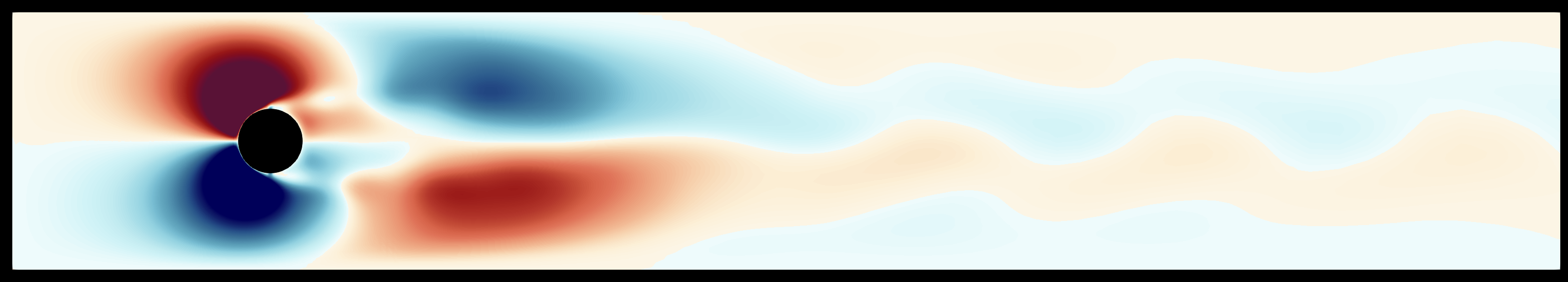}
\put(2,7.5){\textcolor{black}{\scalebox{0.85}{(c)}}}
\end{overpic}%
\begin{overpic}[width=0.049\linewidth,trim={0cm 0cm 0cm 0cm},clip]{figs/cyl_results/scale.png}
\put(17,1){\scalebox{0.57}{-0.2}}
\put(17,94){\scalebox{0.57}{+0.2}}
\end{overpic}
\caption{
Vertical velocity $\phi_v$ fields at $k=499$ (last snapshot) for (a) the uncontrolled flow; (b) the controlled flow with 14 sensors; and (c) the controlled flow with 7 sensors.}
\label{fig:cyl_snaps}
\end{figure}

For the case study with 7 sensors shown in figure \ref{fig:7sensors}, state estimates during the open-loop stage are able to follow the tendencies of the actual states. The same behaviour is seen during the closed-loop control stage, where the NNC is able to significantly attenuate vortex shedding. However, after reaching state space regions close to the natural equilibrium point, high frequency oscillations are introduced by closed-loop dynamics, which makes the NNC respond with small perturbations that hinder convergence. To illustrate the difference between results with different numbers of sensors, figure \ref{fig:cyl_snaps} presents $\phi_v$ contours for the uncontrolled flow, as well as the final snapshot for each controlled case. The levels presented are sufficiently saturated such that small oscillations can be visualized. In both controlled cases, the flow gets considerably closer to the equilibrium point in comparison with the uncontrolled flow, but weaker wake oscillations still propagate, which is more prominent in the case with fewer sensors. Movie 2, provided as supplementary material, plots $\phi_v$ contours evolving with time for each case.

\subsection{Turbulent boundary layer}

The NNs depicted in figures \ref{fig:bl_nnsm}, \ref{fig:bl_nnout} and \ref{fig:bl_nno} are trained by using the first 1500 snapshots (from a total of 1999) from the experimental boundary layer dataset. The same procedure is applied for the super-resolution NN mentioned in \S\ref{sec:met_bl}, which is used to compare the results with the proposed estimator methodology. 

\begin{figure}
\centering
\includegraphics[width=0.98\linewidth]{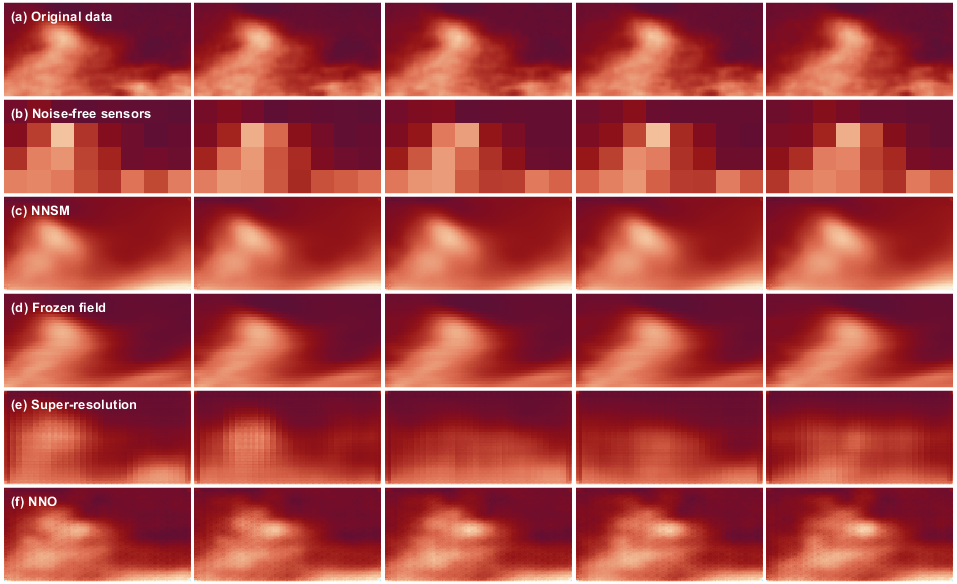}
\begin{overpic}[width=0.97\linewidth]{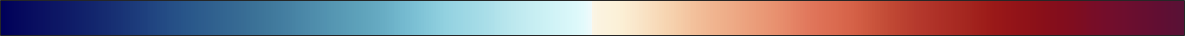}
\put(0.5,0.9){\color{white}\scalebox{.8}{0.30m/s}}
\put(93.2,0.9){\color{white}\scalebox{.8}{4.34m/s}}
\end{overpic}
\caption{Contours of $\phi_u$ from $k=1780$ (leftmost) to $k=1788$ (rightmost), skipping the odd frames. The rows show (a) the original data, (b) the low-resolution sensor data before adding the noise source, (c) the NNSM evolution, (d) the FF model evolution, (e) the direct reconstruction from noisy sensor data, and (f) the NNO estimation using the NNSM and noisy sensor data. These results are also illustrated in more detail in movie 3.}
\label{fig:snap_res_u}
\end{figure}

\begin{figure}
\centering
\includegraphics[width=0.98\linewidth]{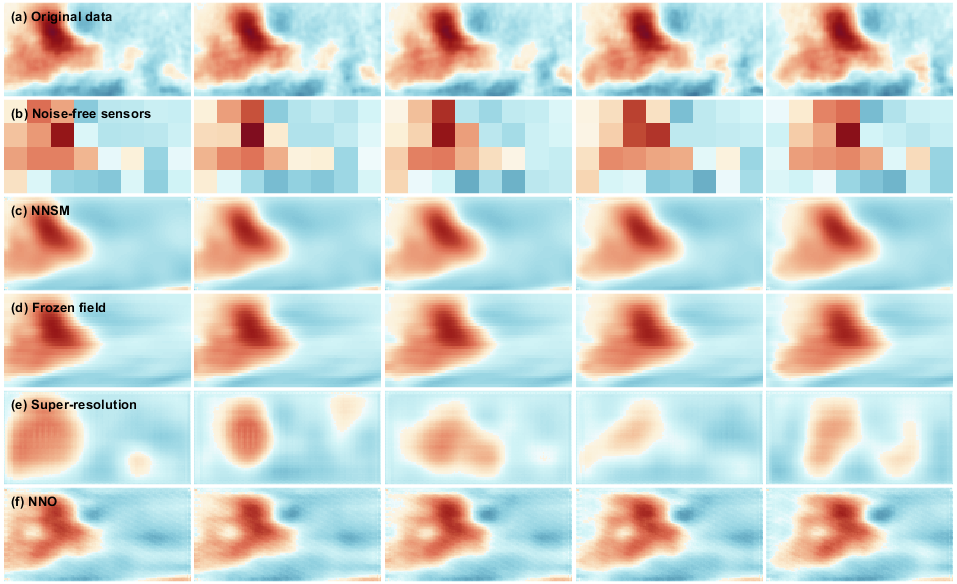}
\begin{overpic}[width=0.97\linewidth]{figs/cbar.png}
\put(0.5,0.9){\color{white}\scalebox{.8}{-0.98m/s}}
\put(93.2,0.9){\color{white}\scalebox{.8}{1.14m/s}}
\end{overpic}
\caption{As in figure \ref{fig:snap_res_u}, but for $\phi_v$.}
\label{fig:snap_res_v}
\end{figure}

Flow snapshots are presented in figures \ref{fig:snap_res_u} and \ref{fig:snap_res_v} for $\phi_u$ and $\phi_v$, respectively. The images are measurements from $k=1780$ to $k=1788$, skipping every odd frame, evolving from left to right. 
For both figures, row (a) presents the original PIV data, while (b) presents a downscaled solution, built by selecting pixel values at the proposed sensor locations. The latter is shown before the addition of measurement noise $\bm{y}_n$. The temporal evolution of the NNSM is shown in row (c), where only the initial conditions 
$\bm{x}[k=0]$
are provided and states are obtained iteratively through $\bm{x}[k+1]=\widetilde{\mathcal{F}}(\bm{x}[k])$. Noticeably, the model tends to smoothen smaller structures, but is able to reproduce the convective characteristics of the flow. Also, for time instants shown, which are far from the initial condition, it is possible to observe that the average flow is shifted, which can be seen as lighter colours occurring in the top-right corner of the images in row (c). This leads to relatively high estimation errors, as will be further discussed. The FF flow evolution is shown in row (d), and as discussed below, the errors seen are smaller than those found for the NNSM, at least under ideal circumstances. We choose the NNSM instead of the FF model for training the NNO to avoid an a priori assumption of the plant characteristics, as well as to show that a certain level of robustness to model imperfections can be expected of the NNO closed-loop estimation. This is a desirable feature, since in real-world control systems plant variations are expected, and a closed-loop controller/observer can often compensate for such variations.

The results of direct reconstructions from noisy sensor data inputs are presented in row (e) of Figs. \ref{fig:snap_res_u} and \ref{fig:snap_res_v}. The super-resolution NN, trained with a structure similar to the NNO --- although without the $\hat{\bm{y}}$ inputs --- is presented here to show the advantages of leveraging sensor data for state estimation. The contours show that the noisy measurements make the reconstruction prone to temporal instability, which can be observed by the intermittent structures that quickly appear and disappear. This is expected since the direct reconstruction saves no memory from past estimations. On the other hand, NNO results illustrated in row (f) provide the smallest error values with improved temporal stability, which is achieved by leveraging both the noisy measurement data and the imperfect NNSM. The evolution of the estimated states through the NNSM, i.e., the prediction step, saves information from previous estimations, thus allowing for rejecting part of the noise introduced to sensors. The noisy low-resolution images, in turn, can be used to reduce errors caused by imperfect predictions, such as strong smoothening and the mean flow offsets. For cases (c), (d) and (f), $\bm{x}[k=0]=\bar{\phi}_u(x,y)$ is used. For better visualization of each case, movie 3 is provided as supplementary material, showing the comparison for the entire time series.

\begin{figure}
\centering
\includegraphics[width=0.98\linewidth,trim={2.8cm .7cm 2cm 0cm},clip]{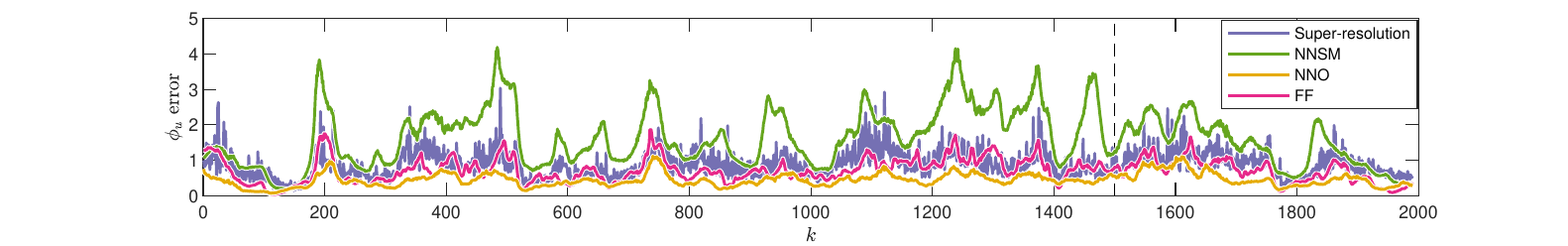}
\includegraphics[width=0.98\linewidth,trim={2.8cm 0cm 2cm .2cm},clip]{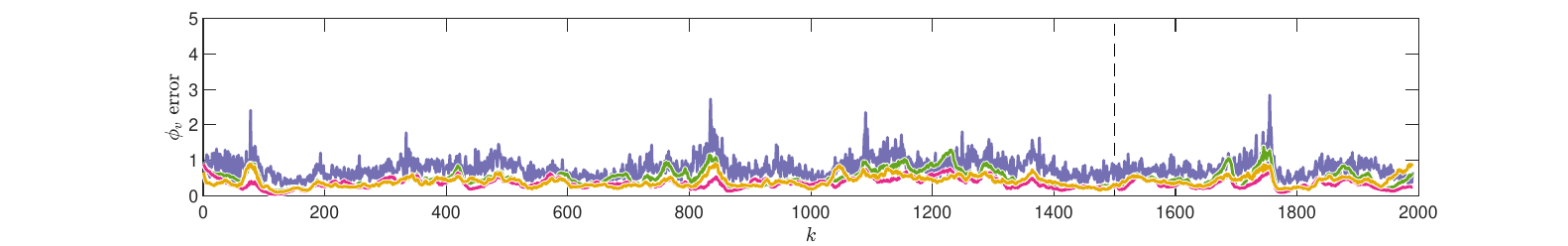}
\caption{
Comparison between the errors of the flow estimation methods shown in figures \ref{fig:snap_noise_u} and \ref{fig:snap_res_v}. Only data to the left of the vertical dashed line is used for training.}
\label{fig:error_comp}
\end{figure}

\begin{figure}
\centering
\includegraphics[width=0.98\linewidth,trim={2.5cm .75cm 2cm 0cm},clip]{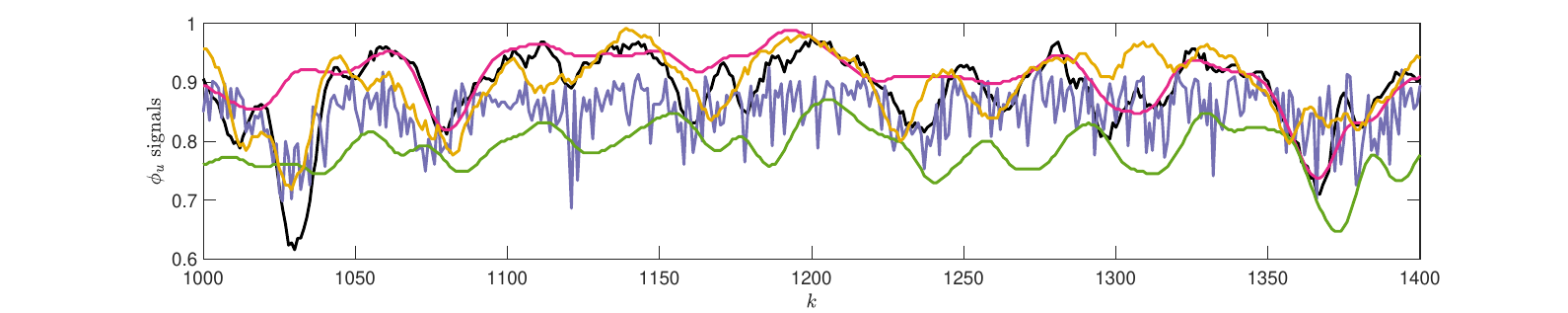}
\includegraphics[width=0.98\linewidth,trim={2.5cm 0cm 2cm .2cm},clip]{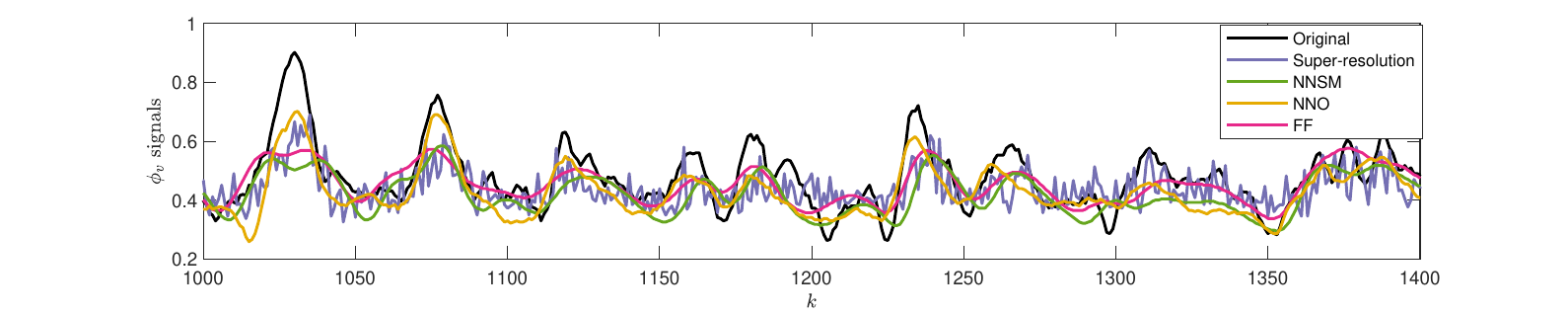}
\caption{Velocities $\phi_u$ and $\phi_v$ computed at halfway up the PIV window height and ay 94\% of the distance between the inflow and the outflow.}
\label{fig:signals_comp}
\end{figure}

To compare results for each case, we propose the squared error metric normalized by the velocity fluctuations
\begin{align}
e_{\phi_u} [k] &= \frac{\|\phi_u[k] - \hat{\phi}_u[k]\|^2}{\|\phi_u[k]-\bar{\phi}_u(x,y)\|^2} \mbox{ ,} \\
e_{\phi_v} [k] &= \frac{\|\phi_v[k] - \hat{\phi}_v[k]\|^2}{\|\phi_v[k]-\bar{\phi}_u(x,y)\|^2} \mbox{ ,}
\end{align}
where the norms are computed by considering each image as a flattened vector. The horizontal and vertical velocity components of the estimated states are represented as $\hat{\phi}_u$ and $\hat{\phi}_v$, respectively. The errors along time are presented in figure \ref{fig:error_comp}. 
The NNSM case uses only the model to propagate the boundary conditions, while the super-resolution NN directly reconstructs the velocity fields from noisy sensor data. While the former presents higher error values (particularly for $\phi_u$), the super-resolution NN presents strong oscillations due to measurement noise. By combining both sensor data and the model, the NNO loop is able to achieve better temporal stability and smaller errors, comparable to the FF model, with the advantage of leveraging output feedback to enable increased robustness in real-world applications, which can be crucial for closed-loop control strategies. 
Figure \ref{fig:signals_comp} presents $\phi_u$ and $\phi_v$ predictions in the range $k=1000$ to $k=1400$, measured at half the PIV window height and at 94\% of the streamwise distance between inflow and outflow. For this probe located near the outflow boundary of the spatial domain, the $\phi_u$ signal reconstructed by the NNSM exhibits a displacement relative to the original flow signal. In addition, both $\phi_u$ and $\phi_v$ obtained from the NNSM display a slight phase shift. The super-resolution approach, in turn, yields compromised results due to the noisy input data. The FF model is able to reproduce part of the original fluctuations but overly smooths both $\phi_u$ and $\phi_v$. The most accurate reconstruction is provided by the NNO, which successfully captures the majority of oscillations in both velocity components. For $\phi_v$, the peaks and phases of the oscillations are resolved with higher fidelity, while for $\phi_u$ the agreement is less precise, but still satisfactory. Although results are not perfect, we must emphasize that this is a convection-dominated problem and small structures may pass through the flow window without being sensed by probes.

\begin{figure}
\centering
\includegraphics[width=0.98\linewidth,trim={0cm 0cm 0cm 0cm},clip]{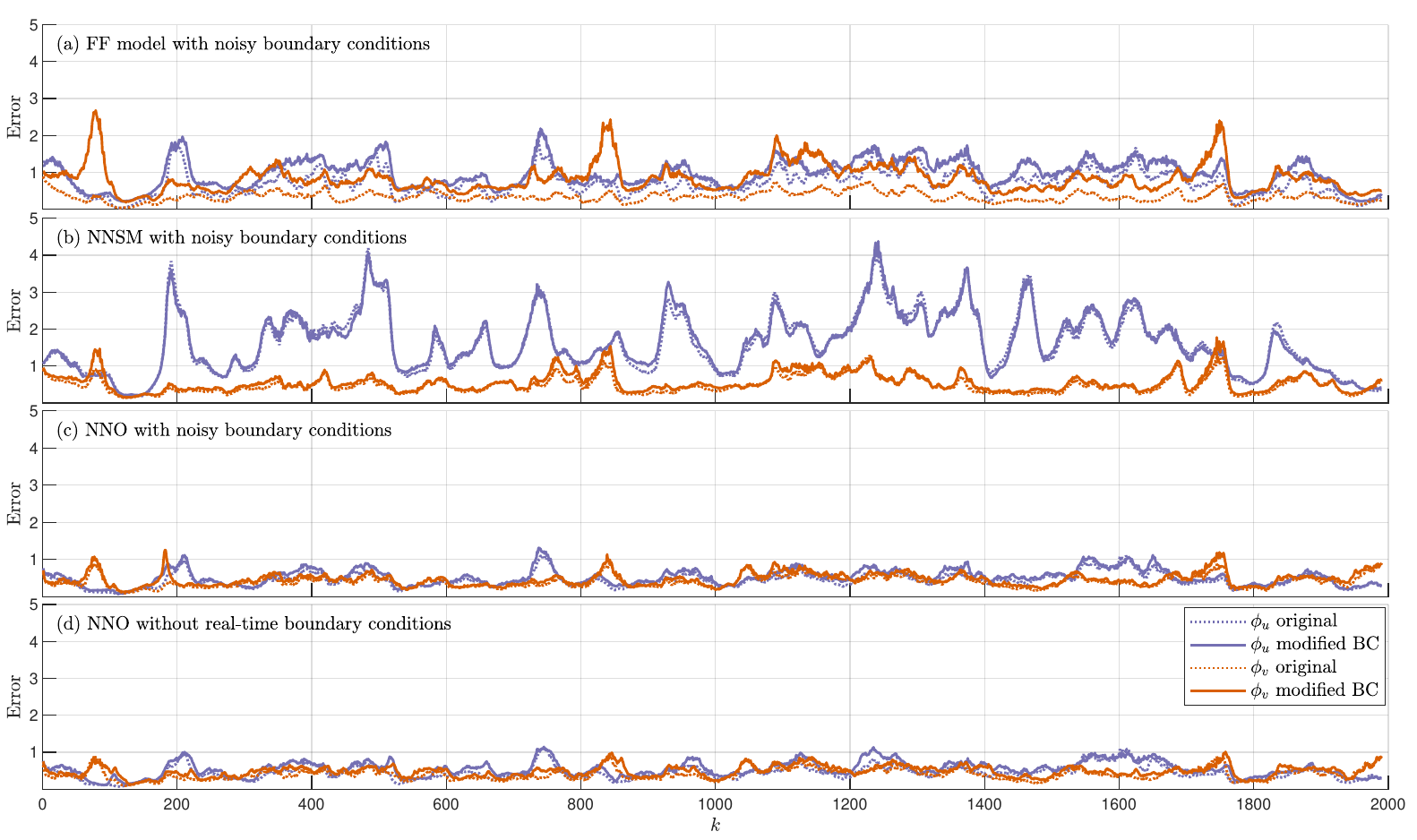}
\caption{Errors before and after the contamination of real-time inlet boundary condition signals with white noise.}
\label{fig:error_allinone}
\end{figure}

\begin{figure}
\centering
\includegraphics[width=0.98\linewidth]{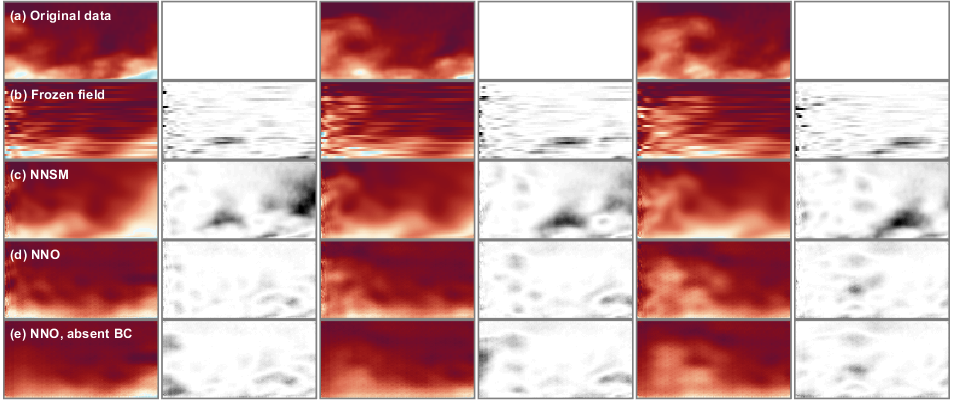}
\begin{overpic}[width=0.975\linewidth]{figs/cbar.png}
\put(0.5,0.9){\color{white}\scalebox{.8}{0.30m/s}}
\put(93.2,0.9){\color{white}\scalebox{.8}{4.34m/s}}
\end{overpic}
\begin{overpic}[width=0.975\linewidth]{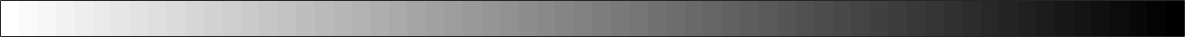}
\put(0.5,0.9){\color{black}\scalebox{.8}{0.00m$^2$/s$^2$}}
\put(91.5,0.9){\color{white}\scalebox{.8}{1.26m$^2$/s$^2$}}
\end{overpic}
\caption{Contours of $\phi_u$ for $k=590$, $600$ and $610$ with respective squared differences. The rows show the (a) original data, (b) the FF model evolution, (c) the NNSM evolution, (d) the NNO estimation with boundary conditions. Noisy boundary conditions are imposed in (b), (c) and (d), while row (e) shows the NNO estimation without real-time boundary conditions. }
\label{fig:snap_noise_u}
\end{figure}

\begin{figure}
\centering
\includegraphics[width=0.98\linewidth]{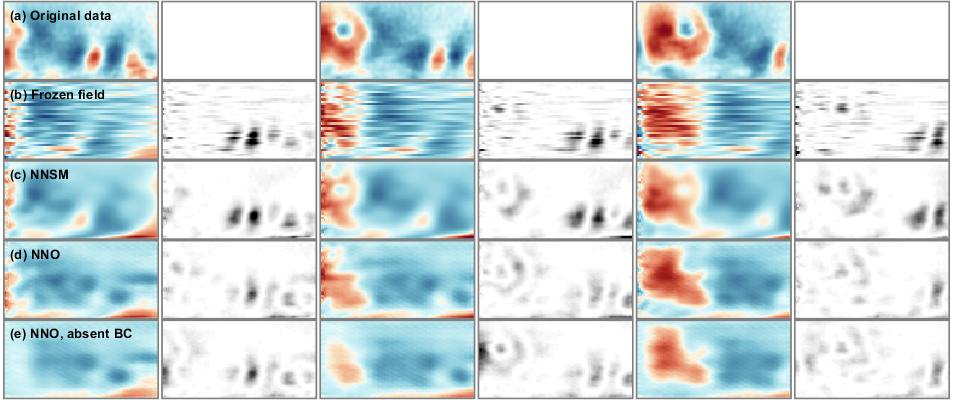}
\begin{overpic}[width=0.975\linewidth]{figs/cbar.png}
\put(0.5,0.9){\color{white}\scalebox{.8}{-0.98m/s}}
\put(93.2,0.9){\color{white}\scalebox{.8}{1.14m/s}}
\end{overpic}
\begin{overpic}[width=0.975\linewidth]{figs/cbar2.png}
\put(0.5,0.9){\color{black}\scalebox{.8}{0.00m$^2$/s$^2$}}
\put(91.5,0.9){\color{white}\scalebox{.8}{0.69m$^2$/s$^2$}}
\end{overpic}
\caption{As in figure \ref{fig:snap_noise_u}, but for $\phi_v$.}
\label{fig:snap_noise_v}
\end{figure}

We additionally test the performance of the estimation/reconstruction systems when random white noise is added to the $\bm{u}$ vector containing the flow boundary conditions. We add the noise source with standard deviation 0.18, the same that was used to contaminate the sensor data. Figure \ref{fig:error_allinone} (a) shows the results in terms of error for the FF case. The curves show that the model provides considerably larger errors when the signal is contaminated with noise, especially for predictions of $\phi_v$. Figure \ref{fig:error_allinone} (b) shows the same curves for the NNSM, which shows more subtle error differences. Here, we use the same NNSM trained without adding noise to $\bm{u}$ in order to remain faithful to the methodology proposed. This means that there could be room for improvement if a noise source with similar characteristics were introduced during training to make the models more robust. 
The ability to reject noise also reflects in lower error variations for the NNO loop, whose results are shown in figure \ref{fig:error_allinone} (c), also exhibiting resilience in the presence of noise while keeping smaller error values by leveraging sensor data. Finally, figure \ref{fig:error_allinone} (d) presents the results for a case where the real-time boundary conditions are unseen by the NNO loop. In this case, the mean flow $\bar{\phi}_u(x=0,y)$ is fed to the NNSM in the NNO loop at every time step. The error increases only marginally, and mostly affects the inflow region, as shown in figures \ref{fig:snap_noise_u} and \ref{fig:snap_noise_v}.
There, the velocity contours for absent and noise-corrupted boundary conditions are shown with the pixel-wise squared difference. Here, instead of showing every consecutive time step, we show a sequence at $k=590$, $k=600$ and $k=610$. As large scale structures enter the domain, more intense errors are seen at the left border when no real-time boundary conditions are provided. In the last snapshot, when the structure is already within the domain and reaches the sensors, the error is reduced along this region. Without the boundary conditions, the NNSM and the FF model are unable to work alone, since no structure is present on the inlet to be transported with the flow. Therefore, no comparison is relevant in these cases. The above results are also illustrated in movie 4, submitted as supplementary material.

\section{Conclusions} \label{sec:conclusions}

We develop and apply a machine learning framework for real-time sensor-based state estimation using dynamic surrogate models of a given plant. The proposed method estimates unsteady flow variables under limited sensing conditions. Unlike conventional approaches that directly reconstruct flow states from sparse measurements, our formulation is inspired by control theory, combining prediction and correction steps. In this context, machine learning serves as a means to extend classical linear approaches, specifically the Luenberger observer in combination with full state feedback control. 
The proposed methodology is tested with three dynamical systems including a modified Kuramoto-Sivashinsky equation, a confined cylinder flow, and a turbulent boundary layer. The two first cases are investigated using numerical simulation models and include flow estimation and closed-loop control, while the third case employs experimental data from PIV measurements, and only considers flow estimation.

The study conducted on the controlled KS equation serves as a proof of concept of the methodology capability to estimate states of discretized partial differential equations. By measuring only a subset of the state space, we demonstrate that the states can be accurately estimated, even in the presence of different types of noise and with coarse sampling intervals. The estimations enable closed-loop control via full state feedback using an NNC, allowing for wave attenuation up to the point where measurement noise dominates the output signals. In contrast to an extended Kalman filter, which requires variable gain updates based on initial error statistics, the optimal estimator obtained through training does not incorporate a variable gain and provides solutions based on the state/output relationships observed during training. The Kalman filter requires an estimate of the initial error covariance to iteratively compute a gain that converges according to measurement noise statistics. In contrast, our approach is trained without any prior knowledge of initial condition errors, and its performance is evaluated solely based on the current estimated and measured outputs. Moreover, the proposed NNO approach does not require local linearizations, as it is trained with the nonlinear models in the loop.


To assess the performance of the proposed NNO in an actual flow problem, we demonstrate its ability to estimate a discretized velocity field in the flow past a cylinder immersed in a plane channel. In this case, 306 state variables, corresponding to the velocity components along the cylinder wake, are estimated using pressure sensors located on the cylinder surface and the channel walls. For both configurations tested, employing either 14 or 7 sensors, the estimated states enabled closed-loop control that attenuated vortex shedding. Although the stabilizing NNC did not fully suppress the instabilities, it achieved a substantial reduction in oscillation amplitudes without requiring retraining of the NNC. The use of only a few wall-mounted pressure sensors highlights the potential of this approach for experimental applications, where direct velocity measurements in regions away from the walls are difficult to obtain.

Aiming to bridge the gap between simulations and experiments in the field of flow control, we additionally tested the training approach with legacy data provided from a PIV of a turbulent boundary layer. By implementing specialized architectures for each of the involved NNs, we achieve promising results, demonstrating that, in specific regions of the state space, it is feasible to estimate the main flow features from low-resolution, noise-corrupted experimental data.
Furthermore, we show that combining such data with an NNSM enhances the estimation accuracy compared to employing either strategy individually, while also providing a degree of robustness to unforeseen phenomena, such as noisy boundary conditions implemented as control inputs. The turbulent flow under consideration is high-order and inherently complex, making precise estimation of all flow variables challenging, if not impossible. 
For certain problems, such as flows exhibiting periodic vortex shedding, the presence of dominant modes facilitates the inference of flow states from a limited set of measurements. However, in control applications, perturbations can reveal stable modes that remain unobserved under unforced conditions.


For problems similar to the present boundary layer, sparse sensor data can lead to ambiguous representations of the size and shape of turbulent structures, and smaller scales may remain unresolved due to 
low-resolution measurements.
Future experimental campaigns, which are beyond the scope of the present work, could assess the performance of feedback control systems driven by such imprecise state estimates. Additional possibilities include the application of the proposed NNO-NNC framework to these experimental settings.
This will require the development of models with improved parameter efficiency to enable real-time implementation, allowing for model embedding into a field programmable gate array (FPGA) system or to a microcontroller. It may also be necessary to tune the models encompassing all NNs employed in the current approach, potentially incorporating observers with internal memory to represent states in a latent space, thereby enabling dynamic output feedback.

\section*{Acknowledgements}
We thank Dr. Fernando Zigunov (Syracuse University) for providing the experimental data and important technical information regarding the boundary layer experiment. We also thank Dr. Fulvio Scarano (TU Delft) and Dr. Meelan Choudhari (NASA Langley Research Center) for their foundational insights that made this work possible.

\section*{Funding}
The authors acknowledge the financial support received from Fundação de Amparo à Pesquisa do Estado de São Paulo, FAPESP, under Grants No. 2013/08293-7, 2021/06448-0 and 2024/21444-9, and from Conselho Nacional de Desenvolvimento Científico e Tecnológico, CNPq, under Grants No. 304320/2024-2 and 444329/2024-2. Financial support from Universidade Estadual de Campinas, through FAEPEX Grant No. 2375/24, is also acknowledged. STMD acknowledges support from the US National Science Foundation award \#2238770.

\section*{Appendix A. Details on legacy neural networks}
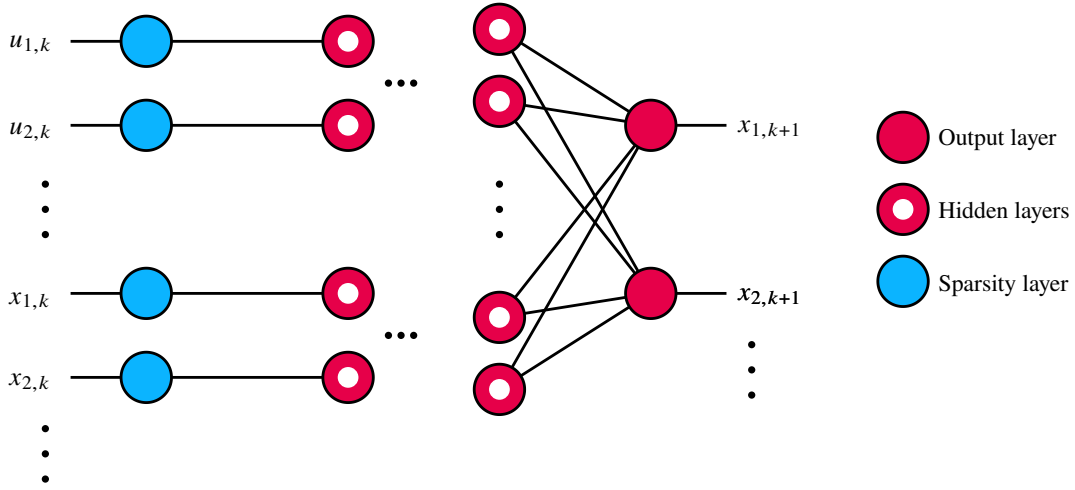
\begin{figure}
\centering
\begin{overpic}[width=.88\linewidth,trim={3.4cm 2cm 0cm 2cm},clip]{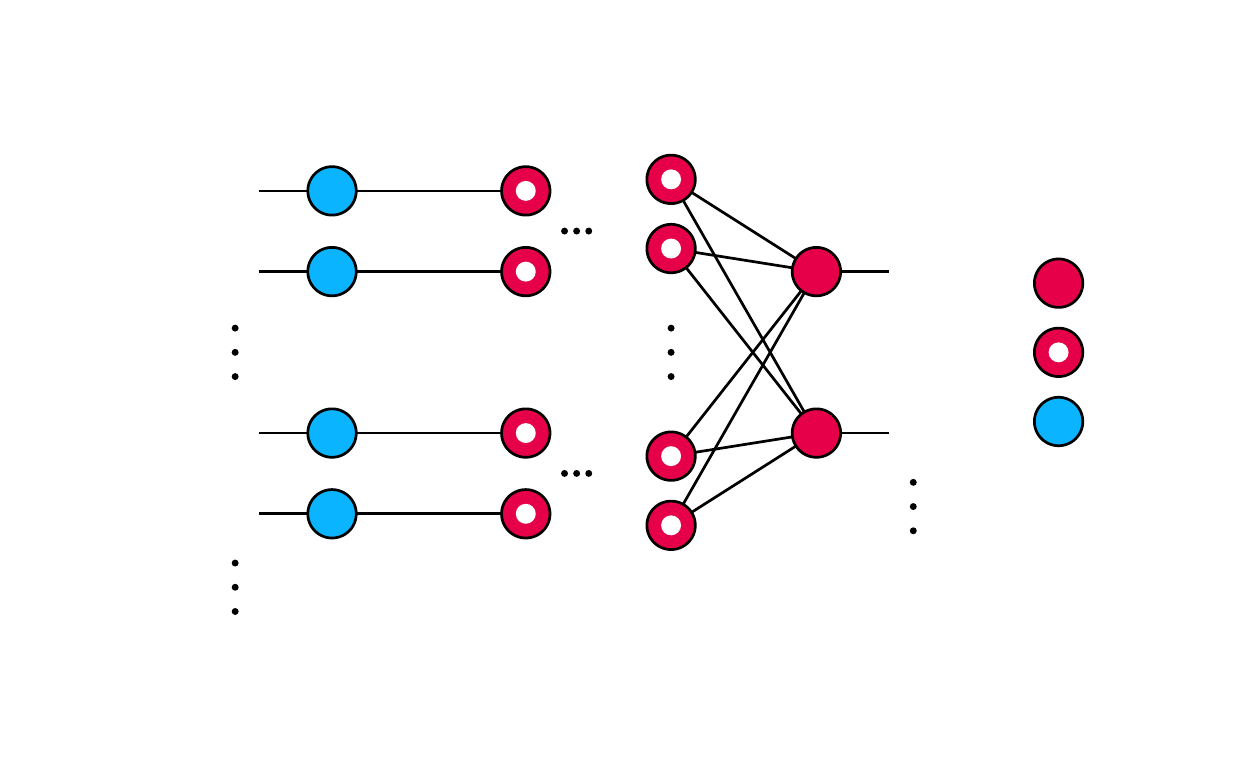}
\put(0,44.5){$u_{1,k}$}
\put(0,36.5){$u_{2,k}$}
\put(0,21.0){$x_{1,k}$}
\put(0,13.5){$x_{2,k}$}
\put(66.7,36.7){$x_{1,k+1}$}
\put(66.7,21.3){$x_{2,k+1}$}
\put(66.7,21.3){$x_{2,k+1}$}
\put(85.0,35.5){\footnotesize{Output layer}}
\put(85.0,28.8){\footnotesize{Hidden layers}}
\put(85.0,22.1){\footnotesize{Sparsity layer}}
\end{overpic}
\caption{Schematic of the NNSM where the red (blue) nodes are related to weights penalized by the L2 (L1) regularization. A ReLU function is employed in the nodes with a white mark, while a linear function is used for the other nodes.}
\label{fig:appa1}
\end{figure}

The NNSM and NNC employed for the KS and the cylinder flow cases studied in this work were trained a priori, and the implementation details are reported by \cite{deda2024neural}. The main structure of the NNSMs consists of fully connected layers, preceded by a direct transfer layer, whose weights are penalized using L1 regularization to eliminate some of the input states from the calculations. Figure \ref{fig:appa1} illustrates this architecture. For the NNC, the structure is similar, but the direct transfer layer is comprised of zeros (at the same locations where zeros are present at the respective trained NNSM) and ones (otherwise). Table \ref{tab:legacy_pars} presents the final number of states that are not excluded in each case, as well as the number of layers and nodes to summarize the complexity of the NNs. For further training details, we refer the reader to the original article. 

\begin{table}
  \begin{center}
\def~{\hphantom{0}}
  \begin{tabular}{lccccc}
    Neural network & Original input states & Final input states & NN number of layers & Layer nodes\\[3pt]
    KS, NNSM & 60 & 60 & 1 & [18]\\
    KS, NNC  & 60 & 60 & 1 & [8]\\
    Cylinder, NNSM        & 306 & 43 & 2 & [100, 80]\\
    Cylinder, NNC         & 306 & 43 & 1 & [8]\\
  \end{tabular}
  \caption{Details of the NNC and NNSM used for de KS and the cylinder flow cases.}
  \label{tab:legacy_pars}
  \end{center}
\end{table}

\section*{Appendix B. Neural network observer details}

The NN output models and NNOs trained for the KS and the cylinder flow cases are simply composed of fully connected layers. Their hyperparameters are summarized in table \ref{tab:ks_nno_pars}. For the cylinder case, the dataset was split in batches. The horizon length used to compute the NNO loss function started with $n_h = 5$ for both cases, being incremented by one every 300 epochs from the 600th epoch for the KS case, and for every 200 epochs from the 200th epoch in the cylinder flow, being capped at the values shown in the table. Adam optimization is chosen, with the learning rates specified in the table.

\begin{table}
  \begin{center}
\def~{\hphantom{0}}
  \begin{tabular}{lcc}
    Hyperparameter & KS cases & Cylinder flow \\[3pt]
    $n_h$ & 5 $\xrightarrow{}$ 18 & 5 $\xrightarrow{}$ 12 \\
    Output model layers & [18] & [100, 80] \\
    Output model learning rate & 0.006 & 0.002 \\
    Output model epochs & 6000 & 3000 \\
    $\lambda_{\widetilde{\mathcal{C}}}$ & $1 \times 10^{-7}$ & $1 \times 10^{-4}$ \\
    NNO Layers & [18, 18] & [100, 80] \\
    NNO learning rate & 0.015 & 0.000333 \\
    NNO epochs & 6000 & 24000 \\
    $\lambda_{\mathcal{L}}$ & $1 \times 10^{-7}$ & $3 \times 10^{-6}$ \\
    $\alpha_v$ & 2 & 2 \\
    $\alpha_x$ & 0.05 & 0.5 \\
    Batch size & --- & 150 \\
    $\bm{x}[k=0]$ & $\phi(x_c)=V$ & Near equilibrium\\
    OL+CL steps & 2000 & 200 \\
    OL steps & 2000 & 200 \\
    CL steps & 300 & 350 \\
  \end{tabular}
  \caption{Hyperparameters of the NN output model and NNO.}
  \label{tab:ks_nno_pars}
  \end{center}
\end{table}

The data for training the NN output model and the NNO are obtained by applying open-loop control perturbations to the solvers, starting from initial conditions close to the equilibrium points of the system. For the KS cases, the initial conditions are set as $\phi(x_c)=V$, while for the cylinder, equilibrium is reached by letting the NNC stabilize the flow. The perturbations are first applied with the pre-trained NNC turned on, assuming full state feedback, since the NNO is not trained yet, configuring a combined open-loop/closed-loop stage (OL+CL). With the stabilizing controller turned on, the perturbations allow for sweeping the state space near the equilibrium point. After that, a stage with only open-loop perturbations is applied to produce data in a broader region of the state space (OL). Finally, the last stage is run with closed-loop control without the open-loop perturbations (CL). The number of time steps for each stage is also shown in table \ref{tab:ks_nno_pars}.

\section*{Appendix C. Neural networks of boundary layer case}
To ensure a better long-term prediction capability of the boundary layer NNSM presented in figure \ref{fig:bl_nnsm}, we train the network weights by unrolling the model response in time within a finite horizon $n_{h\mathrm{s}}$. From the initial dataset containing the initial states $\mathsfbi{X}_0$ for training, we compute $\widetilde{\mathsfbi{X}}_i = \widetilde{\mathcal{F}}(\widetilde{\mathsfbi{X}}_{i-1})$ aiming to minimize the element-wise average of the loss expression
\begin{equation}
    l_{\widetilde{\mathcal{F}}} = \sum_{i=1}^{n_{d\mathrm{s}}}\sum_{k=1}^{n_{h\mathrm{s}}} (\mathsfbi{X}_k - \widetilde{\mathsfbi{X}}_k)  \mbox{ ,}
\end{equation}
where $n_{d\mathrm{s}}$ is the batch number of samples and $\mathsfbi{X}_k$ comes from the experiment. During training, a linear warm-up is applied, where the learning rate grows linearly from 0 to 0.001 during 50 epochs. Table \ref{tab:nnsm_hyperparams} summarizes the hyperparameters for the NNSM. 

\begin{table}
  \begin{center}
\def~{\hphantom{0}}
  \begin{tabular}{lc}
    Hyperparameter & Value \\[3pt]
    Horizon length $n_{h\mathrm{s}}$ & 8 \\
    Batch size $n_{d\mathrm{s}}$ & 64 \\
    Final learning rate & 0.001 \\
    Warm-up epochs & 50\\
    Number of epochs & 1000 \\
    CNN filter size & $3\times3$ \\
  \end{tabular}
  \caption{Hyperparameters of the boundary layer NNSM.}
  \label{tab:nnsm_hyperparams}
  \end{center}
\end{table}

For the output model, the learning rate modulation is performed via discrete decay only. The hyper parameters are shown in table \ref{tab:om_hyperparams}. Finally, the hyperparameters used for the NNO are shown in table \ref{tab:nno_hyperparams}. Similarly to the NNSM, a linear warm-up profile and a discrete exponential decay is applied here. 

\begin{table}
  \begin{center}
\def~{\hphantom{0}}
  \begin{tabular}{lc}
    Hyperparameter & Value \\[3pt]
    Batch size & 128 \\
    Initial learning rate & 0.0015 \\
    Decay rate per step & 0.93 \\
    Epochs for decay & 100 \\
    Number of epochs & 1500 \\
    CNN filter size & $3 \times 3$ \\
    L2 regularization weight $\lambda_{\widetilde{\mathcal{C}}}$ & $1 \times 10^{-6}$ \\
  \end{tabular}
  \caption{Hyperparameters of the boundary layer output model.}
  \label{tab:om_hyperparams}
  \end{center}
\end{table}

\begin{table}
  \begin{center}
\def~{\hphantom{0}}
  \begin{tabular}{lc}
    Hyperparameter & Value \\[3pt]
    Horizon length & 8 \\
    Batch size & 64 \\
    Initial learning rate & 0.001 \\
    Number of epochs & 600 \\
    Warm-up epochs & 100 \\
    L2 regularization weight $\lambda_{\mathcal{L}}$ & $1 \times 10^{-4}$ \\
    Transposed convolution filter size & $3 \times 3$ \\
    Noise standard deviation & 0.18 \\
    States weight $w_x$ & 1 \\
    Compensation weight $w_v$ & 0.1 \\
  \end{tabular}
  \caption{Hyperparameters of the boundary layer NNO.}
  \label{tab:nno_hyperparams}
  \end{center}
\end{table}

\bibliographystyle{unsrt}  
\bibliography{references}  

@article{fukami2019, 
title={Super-resolution reconstruction of turbulent flows with machine learning}, 
volume={870}, DOI={10.1017/jfm.2019.238}, 
journal={Journal of Fluid Mechanics}, 
publisher={Cambridge University Press}, 
author={Fukami, Kai and Fukagata, Koji and Taira, Kunihiko}, 
year={2019}, 
pages={106–120}
}

@article{towne2018, 
  title={Spectral proper orthogonal decomposition and its relationship to dynamic mode decomposition and resolvent analysis}, 
  volume={847}, 
  DOI={10.1017/jfm.2018.283}, 
  journal={Journal of Fluid Mechanics}, 
  publisher={Cambridge University Press}, 
  author={Towne, A. and Schmidt, O. T. and Colonius, T.}, 
  year={2018}, 
  pages={821–867}
  }

@article{Brener,
  title = {Active flow control for drag reduction of a plunging airfoil under deep dynamic stall},
  author = {Ramos, Brener L. O. and Wolf, William R. and Yeh, Chi-An and Taira, Kunihiko},
  journal = {Phys. Rev. Fluids},
  volume = {4},
  issue = {7},
  pages = {074603},
  numpages = {23},
  year = {2019},
  month = {Jul},
  publisher = {American Physical Society}
}

@article{Feitosa2025,
  title = {Control of Deep Dynamic Stall by Duty-Cycle Actuation Informed by Stability Analysis},
  author = {Souza, Lucas F. and Wolf, William R. and Safari, Maryam and Yeh, Chi-An},
  journal = {AIAA Journal},
  numpages = {12},
  year = {2025},
}

@article{morton2018deep,
  title={Deep dynamical modeling and control of unsteady fluid flows},
  author={Morton, Jeremy and Witherden, Freddie D and Jameson, Antony and Kochenderfer, Mykel J},
  journal={arXiv preprint arXiv:1805.07472},
  year={2018}
}

@article{bieker2020deep,
  title={Deep model predictive flow control with limited sensor data and online learning},
  author={Bieker, Katharina and Peitz, Sebastian and Brunton, Steven L and Kutz, J Nathan and Dellnitz, Michael},
  journal={Theoretical and Computational Fluid Dynamics},
  pages={1--15},
  year={2020},
  publisher={Springer Berlin Heidelberg}
}

@article{fan2020reinforcement,
  title={Reinforcement learning for bluff body active flow control in experiments and simulations},
  author={Fan, Dixia and Yang, Liu and Wang, Zhicheng and Triantafyllou, Michael S and Karniadakis, George Em},
  journal={Proceedings of the National Academy of Sciences},
  volume={117},
  number={42},
  pages={26091--26098},
  year={2020},
  publisher={National Acad Sciences}
}

@article{deda2021extremum,
  title={Extremum seeking control applied to airfoil trailing-edge noise suppression},
  author={D{\'e}da, Tarc{\'\i}sio C and Wolf, William R},
  journal={AIAA Journal},
  pages={823--843},
  volume={60},
  year={2022},
  publisher={American Institute of Aeronautics and Astronautics}
}

@article{ren2021applying,
  title={Applying deep reinforcement learning to active flow control in weakly turbulent conditions},
  author={Ren, Feng and Rabault, Jean and Tang, Hui},
  journal={Physics of Fluids},
  volume={33},
  number={3},
  pages={037121},
  year={2021},
  publisher={AIP Publishing LLC}
}

@article{gadelhak2001future,
  title={Flow Control: The Future},
  author={Gad-el-Hak, Mohamed},
  journal={Journal of Aircraft},
  volume={38},
  pages={402--418},
  year={2001},
}

@article{brunton2015closed,
  title={Closed-loop turbulence control: progress and challenges},
  author={Brunton, Steven L and Noack, Bernd R},
  journal={Applied Mechanics Reviews},
  volume={67},
  number={5},
  pages={050801},
  year={2015},
  publisher={American Society of Mechanical Engineers}
}

@article{choi1994active,
  title={Active turbulence control for drag reduction in
wall-bounded flows},
  author={Choi, Haecheon and Moin, Parviz and Kim, John},
  journal={Journal of Fluid Mechanics},
  volume={262},
  pages={75--110},
  year={1994},
}

@article{brunton2020machine,
  title={Machine learning for fluid mechanics},
  author={Brunton, Steven L and Noack, Bernd R and Koumoutsakos, Petros},
  journal={Annual Review of Fluid Mechanics},
  volume={52},
  pages={477--508},
  year={2020},
  publisher={Annual Reviews}
}

@article{rabault2019artificial,
  title={Artificial neural networks trained through deep reinforcement learning discover control strategies for active flow control},
  author={Rabault, Jean and Kuchta, Miroslav and Jensen, Atle and R{\'e}glade, Ulysse and Cerardi, Nicolas},
  journal={Journal of fluid mechanics},
  volume={865},
  pages={281--302},
  year={2019},
  publisher={Cambridge University Press}
}

@article{deda2023backpropagation,
  title={Backpropagation of neural network dynamical models applied to flow control},
  author={D{\'e}da, Tarc{\'\i}sio and Wolf, William R and Dawson, Scott T M},
  journal={Theoretical and Computational Fluid Dynamics},
  pages={1--25},
  year={2023},
  publisher={Springer}
}

@article{castellanos2022machine,
  title={Machine-learning flow control with few sensor feedback and measurement noise},
  author={Castellanos, R and Cornejo Maceda, GY and De La Fuente, I and Noack, BR and Ianiro, A and Discetti, S},
  journal={Physics of Fluids},
  volume={34},
  number={4},
  year={2022},
  publisher={AIP Publishing}
}

@article{guastoni2023deep,
  title={Deep reinforcement learning for turbulent drag reduction in channel flows},
  author={Guastoni, Luca and Rabault, Jean and Schlatter, Philipp and Azizpour, Hossein and Vinuesa, Ricardo},
  journal={The European Physical Journal E},
  volume={46},
  number={4},
  pages={27},
  year={2023},
  publisher={Springer}
}

@article{sonoda2023reinforcement,
  title={Reinforcement learning of control strategies for reducing skin friction drag in a fully developed turbulent channel flow},
  author={Sonoda, Takahiro and Liu, Zhuchen and Itoh, Toshitaka and Hasegawa, Yosuke},
  journal={Journal of Fluid Mechanics},
  volume={960},
  pages={A30},
  year={2023},
  publisher={Cambridge University Press}
}

@article{zigunov2023multiaxis,
  title={Multiaxis Shock Vectoring Control of Overexpanded Supersonic Jet Using a Genetic Algorithm},
  author={Zigunov, Fernando and Song, MyungJun and Sellappan, Prabu and Alvi, Farrukh S},
  journal={Journal of Propulsion and Power},
  volume={39},
  number={2},
  pages={249--257},
  year={2023},
  publisher={American Institute of Aeronautics and Astronautics}
}

@article{zigunov2022bluff,
  title={A bluff body flow control experiment with distributed actuation and genetic algorithm-based optimization},
  author={Zigunov, Fernando and Sellappan, Prabu and Alvi, Farrukh},
  journal={Experiments in Fluids},
  volume={63},
  number={1},
  pages={23},
  year={2022},
  publisher={Springer}
}

@article{li2022reinforcement,
  title={Reinforcement-learning-based control of confined cylinder wakes with stability analyses},
  author={Li, Jichao and Zhang, Mengqi},
  journal={Journal of Fluid Mechanics},
  volume={932},
  pages={A44},
  year={2022},
  publisher={Cambridge University Press}
}

@article{willcox2006unsteady,
  title={Unsteady flow sensing and estimation via the gappy proper orthogonal decomposition},
  author={Willcox, Karen},
  journal={Computers \& fluids},
  volume={35},
  number={2},
  pages={208--226},
  year={2006},
  publisher={Elsevier}
}

@article{manohar2018data,
  title={Data-driven sparse sensor placement for reconstruction: Demonstrating the benefits of exploiting known patterns},
  author={Manohar, Krithika and Brunton, Bingni W and Kutz, J Nathan and Brunton, Steven L},
  journal={IEEE Control Systems Magazine},
  volume={38},
  number={3},
  pages={63--86},
  year={2018},
  publisher={IEEE}
}

@article{sashittal2021data,
  title={Data-driven sensor placement for fluid flows},
  author={Sashittal, Palash and Bodony, Daniel J},
  journal={Theoretical and Computational Fluid Dynamics},
  volume={35},
  number={5},
  pages={709--729},
  year={2021},
  publisher={Springer}
}

@article{graff2023information,
  title={Information-Based Sensor Placement for Data-Driven Estimation of Unsteady Flows},
  author={Graff, John and Medina, Albert and Lagor, Francis},
  journal={arXiv preprint arXiv:2303.12260},
  year={2023}
}

@article{williams2022data,
  title={Data-driven sensor placement with shallow decoder networks},
  author={Williams, Jan and Zahn, Olivia and Kutz, J Nathan},
  journal={arXiv preprint arXiv:2202.05330},
  year={2022}
}

@article{raibaudo2020machine,
  title={Machine learning strategies applied to the control of a fluidic pinball},
  author={Raibaudo, Cedric and Zhong, Peng and Noack, Bernd R and Martinuzzi, Robert John},
  journal={Physics of Fluids},
  volume={32},
  number={1},
  year={2020},
  publisher={AIP Publishing}
}

@article{varela2022deep,
  title={Deep reinforcement learning for flow control exploits different physics for increasing {R}eynolds number regimes},
  author={Varela, Pau and Su{\'a}rez, Pol and Alc{\'a}ntara-{\'A}vila, Francisco and Mir{\'o}, Arnau and Rabault, Jean and Font, Bernat and Garc{\'\i}a-Cuevas, Luis Miguel and Lehmkuhl, Oriol and Vinuesa, Ricardo},
  journal={Actuators},
  volume={11},
  number={12},
  pages={359},
  year={2022},
  publisher={MDPI}
}

@article{paris2021robust,
  title={Robust flow control and optimal sensor placement using deep reinforcement learning},
  author={Paris, Romain and Beneddine, Samir and Dandois, Julien},
  journal={Journal of Fluid Mechanics},
  volume={913},
  pages={A25},
  year={2021},
  publisher={Cambridge University Press}
}

@article{deda2024neural,
  title={Neural networks in feedback for flow analysis and control},
  author={D{\'e}da, Tarc{\'\i}sio C and Wolf, William R and Dawson, Scott TM},
  journal={Physical Review Fluids},
  volume={9},
  number={6},
  pages={063904},
  year={2024},
  publisher={APS}
}

@inproceedings{yadaiah2006neural,
  title={Neural network based state estimation of dynamical systems},
  author={Yadaiah, Narri and Sowmya, G},
  booktitle={The 2006 IEEE international joint conference on neural network proceedings},
  pages={1042--1049},
  year={2006},
  organization={IEEE}
}

@book{zigunov2020detailed,
  title={Detailed flow field study of an upswept cylinder wake and experimental optimization using active flow control},
  author={Zigunov, Fernando},
  year={2020},
  publisher={The Florida State University}
}

@article{zigunov2024one,
  title={One-shot omnidirectional pressure integration through matrix inversion},
  author={Zigunov, Fernando and Charonko, John J},
  journal={Measurement Science and Technology},
  volume={35},
  number={12},
  pages={125301},
  year={2024},
  publisher={IOP Publishing}
}

@article{audiffred2023experimental,
  title={Experimental control of Tollmien-Schlichting waves using the Wiener-Hopf formalism},
  author={Audiffred, Diego BS and Cavalieri, Andr{\'e} VG and Brito, Pedro PC and Martini, Eduardo},
  journal={Physical Review Fluids},
  volume={8},
  number={7},
  pages={073902},
  year={2023},
  publisher={APS}
}

@article{visbal2018control,
  title={Exploration of High-Frequency Control of DynamicStall Using Large-Eddy Simulations},
  author={Visbal, Miguel R. and Benton, Stuart I.},
  journal={AIAA Journal},
  volume={56},
  number={8},
  pages={2974--2991},
  year={2018},
}

@article{morimoto2022generalization,
  title={Generalization techniques of neural networks for fluid flow estimation},
  author={Morimoto, Masaki and Fukami, Kai and Zhang, Kai and Fukagata, Koji},
  journal={Neural Computing and Applications},
  volume={34},
  number={5},
  pages={3647--3669},
  year={2022},
  publisher={Springer}
}

@article{nakamura2022robust,
  title={Robust training approach of neural networks for fluid flow state estimations},
  author={Nakamura, Taichi and Fukagata, Koji},
  journal={International Journal of Heat and Fluid Flow},
  volume={96},
  pages={108997},
  year={2022},
  publisher={Elsevier}
}

@article{amaral2021resolvent,
  title={Resolvent-based estimation of turbulent channel flow using wall measurements},
  author={Amaral, Filipe R and Cavalieri, Andr{\'e} VG and Martini, Eduardo and Jordan, Peter and Towne, Aaron},
  journal={Journal of Fluid Mechanics},
  volume={927},
  pages={A17},
  year={2021},
  publisher={Cambridge University Press}
}

@article{martini2020resolvent,
  title={Resolvent-based optimal estimation of transitional and turbulent flows},
  author={Martini, Eduardo and Cavalieri, Andr{\'e} VG and Jordan, Peter and Towne, Aaron and Lesshafft, Lutz},
  journal={Journal of Fluid Mechanics},
  volume={900},
  pages={A2},
  year={2020},
  publisher={Cambridge University Press}
}

@book{friedland2012control,
  title={Control system design: an introduction to state-space methods},
  author={Friedland, Bernard},
  year={2012},
  publisher={Courier Corporation}
}

@article{luenberger2007observing,
  title={Observing the state of a linear system},
  author={Luenberger, David G},
  journal={IEEE transactions on military electronics},
  volume={8},
  number={2},
  pages={74--80},
  year={2007},
  publisher={IEEE}
}

@inproceedings{ramos2020numerical,
  title={Numerical design of Luenberger observers for nonlinear systems},
  author={Ramos, Louise da C and Di Meglio, Florent and Morgenthaler, Valery and da Silva, Lu{\'\i}s F Figueira and Bernard, Pauline},
  booktitle={2020 59th IEEE Conference on Decision and Control (CDC)},
  pages={5435--5442},
  year={2020},
  organization={IEEE}
}

@article{zeitz1987extended,
  title={The extended Luenberger observer for nonlinear systems},
  author={Zeitz, Michael},
  journal={Systems \& Control Letters},
  volume={9},
  number={2},
  pages={149--156},
  year={1987},
  publisher={Elsevier}
}

@article{afri2016state,
  title={State and parameter estimation: A nonlinear Luenberger observer approach},
  author={Afri, Chouaib and Andrieu, Vincent and Bako, Laurent and Dufour, Pascal},
  journal={IEEE Transactions on Automatic Control},
  volume={62},
  number={2},
  pages={973--980},
  year={2016},
  publisher={IEEE}
}

@article{fukami2023super,
  title={Super-resolution analysis via machine learning: a survey for fluid flows},
  author={Fukami, Kai and Fukagata, Koji and Taira, Kunihiko},
  journal={arXiv preprint arXiv:2301.10937},
  year={2023}
}

@article{page2024super,
  title={Super-resolution with dynamics in the loss},
  author={Page, Jacob},
  journal={arXiv preprint arXiv:2410.20884},
  year={2024}
}

@article{zhong2023sparse,
  title={Sparse sensor reconstruction of vortex-impinged airfoil wake with machine learning},
  author={Zhong, Yonghong and Fukami, Kai and An, Byungjin and Taira, Kunihiko},
  journal={Theoretical and Computational Fluid Dynamics},
  volume={37},
  number={2},
  pages={269--287},
  year={2023},
  publisher={Springer}
}

@article{yeh2021network,
  title={Network broadcast analysis and control of turbulent flows},
  author={Yeh, Chi-An and Meena, Muralikrishnan Gopalakrishnan and Taira, Kunihiko},
  journal={Journal of Fluid Mechanics},
  volume={910},
  pages={A15},
  year={2021},
  publisher={Cambridge University Press}
}

@article{brunton2016sparse,
  title={Sparse identification of nonlinear dynamics with control (SINDYc)},
  author={Brunton, Steven L and Proctor, Joshua L and Kutz, J Nathan},
  journal={IFAC-PapersOnLine},
  volume={49},
  number={18},
  pages={710--715},
  year={2016},
  publisher={Elsevier}
}

@inproceedings{payne2024co,
  title={Co-Flow Jet Design Optimization},
  author={Payne, Nickolas E and Zhang, Yang and Cattafesta, Louis N},
  booktitle={AIAA AVIATION FORUM AND ASCEND 2024},
  pages={3631},
  year={2024}
}

@article{taira2022network,
  title={Network-based analysis of fluid flows: Progress and outlook},
  author={Taira, Kunihiko and Nair, Aditya G},
  journal={Progress in Aerospace Sciences},
  volume={131},
  pages={100823},
  year={2022},
  publisher={Elsevier}
}

@article{park2020machine,
  title={Machine-learning-based feedback control for drag reduction in a turbulent channel flow},
  author={Park, Jonghwan and Choi, Haecheon},
  journal={Journal of Fluid Mechanics},
  volume={904},
  pages={A24},
  year={2020},
  publisher={Cambridge University Press}
}

@article{miotto2025,
  title={Pressure field reconstruction with {SIREN}: A mesh-free approach for image velocimetry in complex noisy environments},
  author={Miotto, Renato and Wolf, William and Zigunov, Fernando},
  journal={Experiments in Fluids},
  volume={66},
  pages={151},
  year={2025},
  publisher={Springer Nature}
}

\end{document}